\documentclass[prd, showpacs,preprint, preprintnumbers,amsmath,amssymb,
superscriptaddress,nofootinbib]{revtex4-1}
\usepackage{booktabs}
\usepackage{epsfig}
\usepackage{graphicx}
\usepackage{color}
\usepackage{shuffle}
\usepackage{hyperref}
\usepackage{tikz-feynman}
\usepackage{wrapfig}
\usepackage{cases}
\usepackage[utf8]{inputenc}

\usepackage{stackengine}
\stackMath
\usepackage{graphicx}

\def\sect#1{Sec.~{\ref{#1}}}

\newtheorem{claim}{Claim}

\newtheorem{conjecture}{Conjecture}

\AtBeginDocument{
\heavyrulewidth=.08em
\lightrulewidth=.05em
\cmidrulewidth=.03em
\belowrulesep=.65ex
\belowbottomsep=0pt
\aboverulesep=.4ex
\abovetopsep=0pt
\cmidrulesep=\doublerulesep
\cmidrulekern=.5em
\defaultaddspace=.5em
}

\begin{document}
\title{UV considerations on scattering amplitudes in a web of theories}
\author{John Joseph M. Carrasco}
\affiliation{Department of Physics \& Astronomy, Northwestern University, Evanston, Illinois 60208, USA}
\affiliation{Institut de Physique Theorique, Universite Paris Saclay, CEA, CNRS, F-91191 Gif-sur-Yvette, France}
\author{Laurentiu Rodina}
\affiliation{Institut de Physique Theorique, Universite Paris Saclay, CEA, CNRS, F-91191 Gif-sur-Yvette, France}
\date{\today}
\begin{abstract} 
The scattering predictions of a web of theories including Yang-Mills (YM), gravity, bi-adjoint scalar, the non-linear sigma model (NLSM), Dirac-Born-Infeld-Volkov-Akulov (DBI-VA) and the special Galileon (sGal) form a class of special objects with two fascinating properties: they are related by the double-copy procedure, and they can be defined purely by on-shell constraints. We expand on both of these properties. First we show that NLSM tree-level amplitudes are fully determined by imposing color-dual structure together with cyclic invariance and locality. We then consider how hard-scaling can be used to constrain the predictions of these theories, as opposed to the usual soft-scaling.  We probe the UV by generalizing the familiar BCFW shift off-shell to a novel single hard limit. We show that UV scalings are sufficient to fully constrain: 1.~Bi-adjoint doubly-ordered amplitudes, assuming locality; 2.~NLSM and BI, assuming locality and unitarity; 3.~special Galileon, assuming locality, unitarity, and a UV bound for the general Galileon vertex.  We see how potentially distinct aspects of this UV behavior can be understood and unified via double-copy relations.  Surprisingly, we find evidence that assuming unitarity for these theories may not be necessary, and can emerge via UV considerations and locality alone. These results complete the observations that, like IR considerations, UV scaling is sufficient to fully constrain a wide range of tree-level amplitudes, for both gauge, gravity, and effective field theories.
\end{abstract}
\preprint{NUHEP-TH/19-10}
\maketitle

\tableofcontents
\newpage
\section{Introduction}

 On-shell ideas and methods have transformed our approach to understanding perturbative predictions in relativistic quantum field theories. Doing so has exposed structure and relations between field theories from our most formal UV completions in string theory to the grittiest phenomenological theories living well in the IR, features completely hidden by off-shell Lagrangian formulations. Two such discoveries stand out and will be explored in this article:
\begin{itemize} 
\item Various scattering amplitudes may be determined uniquely by on-shell principles
\item These special scattering amplitudes form an intricate web of relations
\end{itemize}
\subsection{On-shell consistency and uniqueness}

While it is a central tenet of quantum field theory that  through Lagrangians symmetries determine an overwhelming majority of our known theories, it was only recently understood that some symmetries are even more constraining directly at the level of on-shell scattering amplitudes, which can avoid much of complicated and redundant machinery of off-shell descriptions. 

For instance, it is common knowledge that gauge invariance fixes the Lagrangian of gauge theories (like QED, Yang-Mills, but also General Relativity), but in fact gauge invariance and locality alone are sufficient to fully fix amplitudes in YM and GR, with unitarity emerging as a consequence \cite{Arkani-Hamed:2016rak,RodinaGaugeInv}. Similarly, effective field theories (EFT's), long known to satisfy symmetries related to IR properties \cite{Susskind:1970gf,Osborn:1969ku,AdlerZero,Ellis:1970nt,ArkaniHamed2008gz}, were just recently seen to follow directly from the on-shell Adler zero condition \cite{Cheung:2014dqa,Cheung:2016drk,Arkani-Hamed:2016rak,RodinaGaugeInv}. 

Yet interesting soft properties are not unique to EFT's: virtually all theories posses soft theorems, another very well known fact, newly re-discovered and explored in many unexpected contexts \cite{Cachazo:2014fwa,Strominger:2014pwa,He2014laa,Hawking:2016msc}. In \cite{RodinaSoft} the power of the IR was understood in a unified way: soft theorems are sufficient to fully constrain a large variety of theories, including both gauge and effective field theories. This lead to the surprising conclusion that the IR contains all the information needed to rebuild the amplitude - in other words, the IR somehow also knows about the UV pieces of amplitudes.

Closer to the UV probing of amplitudes, an orthogonal, and very practical approach to the on-shell program has been the BCFW recursion \cite{BCFW}, which represented a milestone in the conceptual understanding and technical calculation of scattering amplitudes.  Leveraging insight into the unitarity construction at loop-level with Cauchy's theorem and amplitude UV behavior, it formed a concrete realization that for a special set of {\em constructable} theories, including Yang-Mills and Gravity \cite{Benincasa:2007xk,McGady:2013sga}, only minimal on-shell data -- the three point amplitude -- for these theories was required to completely specify all order predictions.  This is in stark contrast with the requirement for higher-order contact terms in the action (infinitely many for GR), but which only exist to ensure gauge invariance.   Effective field theories on the other hand in general have no such structure -- without fundamental symmetries constraining higher orders, every contact term can have an arbitrary coefficient and so lower-multiplicity amplitudes are not enough to determine the higher-point ones. This is manifested directly in the presence of poles at infinity which obstruct the recursion, a reflection of the bad UV behavior of EFT's. However, the special theories under consideration do satisfy a symmetry which fixes the higher-point contact terms: the Adler zero, which in this sense can be regarded as a ``gauge symmetry'' for these scalar theories. Inspired by this observation, the BCFW recursion was extended to EFT's, by modifying the shifts as to include the Adler zero property \cite{Kampf:2013vha,Kampf:2012fn,Cheung:2015cba,Luo:2015tat,Cachazo2016njl,Low:2019ynd,Elvang:2018dco}.

Whether good or bad, the UV scaling is of central importance, but has often been viewed as an annoying obstacle which in some cases can be miraculously  removed, thus allowing the recursion and computations to take place. A closer analysis of the scaling reveals that, like for gauge invariance or the Adler zero, numerous cancellations take place, suggesting that only very special objects can have this property, which may ultimately be related to a symmetry \cite{ArkaniHamed:2008yf}. And indeed, in \cite{Rodina:2016mbk} it was argued that YM and GR tree-level amplitudes (along with their gauge invariance and unitarity) do in fact follow from locality and an improved UV scaling. But the UV scaling of EFT's, while not as good as that of gauge theories, is still highly improved over naive power counting based on Feynman rules. 

Therefore, in this article we explore the UV properties of various tree-level EFT's, probed by two different deformations: a two line BCFW shift and a single hard limit, as opposed to the usual single soft limit. We will find that NLSM, BI, Gal and sGal are all fixed by locality and demanding some particular large $z$-scaling, bringing EFT UV considerations on the same footing as those of gauge theories. This is particularly surprising since it has long been thought that only IR information can be used to constrain EFT's, since they themselves live in the IR. And perhaps most surprisingly of all, we find that even bi-adjoint scalar amplitudes are fixed by such UV considerations. Since these amplitudes have trivial numerators, this example directly demonstrates that UV scaling is somehow probing unitarity.

We are then left with a puzzling fact, which yet lacks a complete explanation: both the IR (as expressed through the soft theorems) and the UV (probed through hard limits) are sufficient to fully fix a wide range of theories. 
\subsection{A color-kinematic web of amplitude relations}

The other structure that appears to have a certain amount of ubiquity in S-matrix predictions, and relates theories discussed here, is the so called color-dual double-copy structure, originally realized in Yang-Mills and its relation to gravity by Bern, Johansson and one of the current authors (BCJ)~\cite{BCJ,BCJLoop}. This structure allows many amplitudes to be expressed as a generalized product between different building blocks, providing a purely field theoretic understanding and generalization of the celebrated KLT formula \cite{KLT}:
\begin{equation}
\textrm{Gravity}=\textrm{Yang-Mills} \otimes \textrm{Yang-Mills}
\end{equation}

 At the heart of this structure lies the color-kinematic duality, which schematically states that given an amplitude expressed in a color ordered amplitude basis, for example a tree-level YM amplitude:
\begin{align}
\mathcal{A}^{\textrm{YM}}=\sum_{\sigma \in S_{n-2}} c_\sigma A^\textrm{YM}(1,\sigma,n)  \, ,
\end{align}
there exist kinematic functions $n^{\textrm{YM}}_\sigma$ that satisfy the same algebra as the color factors $c_\sigma$, and can replace them, producing the GR amplitude:
\begin{align}
\mathcal{A}^{\textrm{GR}}=\sum_{\sigma \in S_{n-2}} n^{\textrm{YM}}_\sigma A^{\textrm{YM}}(1,\sigma,n) \, .
\end{align}
The story goes deeper, as there exist other similar functions $n^{\textrm{NLSM}}_\sigma$, which instead lead to an expression for the BI amplitudes:
\begin{align}
\mathcal{A}^{\textrm{BI}}=\sum_{\sigma \in S_{n-2}} n^{\textrm{NLSM}}_\sigma A^{\textrm{YM}}(1,\sigma,n) \, .
\end{align}
This structure is what leads to the KLT factorization for adjoint-compatible amplitudes, as well as the scattering equations allowing for the CHY expression of tree-level amplitudes \cite{Cachazo:2013gna,Cachazo:2013hca,Cachazo:2013iea,Cachazo2014nsa,Cachazo2014xea,He:2016iqi}.  The web is even more tangled, as transmutation operators can directly transform some amplitudes into others \cite{CheungUnifyingRelations,Cheung:2017yef}.  Perhaps more intriguing even than such tree-level relations, these kinematic functions $n$ have a local interpretation relevant to a graph representation of the amplitudes -- theory specific dressings dual to $f^{abc}$ color-weights, over scalar propagators.  In essence, these functions have the same job as color-charges weights, locally dressing graphs with some kinematic function relevant to the building blocks of the theory at hand.  This seamlessly generalizes to the multi-loop corrections at the integrand level.   With just a small set of these color-charge like weights, $c$, $n^{\textrm{YM}}$, $n^{\textrm{NLSM}}$ obeying the same algebraic relations (namely Jacobi and antisymmetry about vertices) one can build full multi-loop amplitudes for this family of theories:
\begin{equation}
{\cal A}^{(L)}_n \propto \sum_{i}\, 
 \int \frac{d^{LD}\ell}{(2\pi)^{LD}} \frac{1}{S_i}\frac{m_{i} \widetilde{m}_i}{D_i}  \, ,
\end{equation}
where various theories are given by the  choices of ($m,\widetilde{m}$) given in Table~\ref{JacFactors}. In these cases of an adjoint double-copy, the sums run over all distinct $L$-loop $m$-point cubic diagrams,  the $S_i$ represent the symmetry factors of the graphs, and the $D_i$ are massless scalar propagators relevant to each graph.

\begin{table*}
\begin{center}
\caption{Factorization to adjoint color-dual graph numerators for this double-copy web of theories. }
\label{JacFactors}
\begin{tabular}{ c|c|c} 
 \toprule
Theory &  $m$ & $\tilde{m}$ \\
\midrule
~~Bi-Adjoint $\phi^3$~~&~~$c(f^{abc})$  [color] \, ~~&~~$\tilde{c}(\tilde{f}^{abc})$  (color)~~\\ 
~~Yang-Mills & $c(f^{abc})$ [color]~~&~~$n^{\textrm{YM}}(k,  \epsilon)$  [vector]~~\\ 
~~Gravity + Axion + Dilaton~~&~~$n^{\textrm{YM}}(k ,\epsilon)$ [vector]~~&~~$n^{\textrm{YM}}(k, \epsilon)$ [vector]~~\\ 
~~NLSM~~&~~$c(f^{abc})$ [flavor/color]~~&~~$n^{\textrm{NLSM}}(k)$ [scalar]~~\\ 
~~Born-Infeld~~&~~$n^{\textrm{YM}}(k, \epsilon)$ [vector]~~&~~$n^{\textrm{NLSM}}(k)$ [scalar]~~\\ 
~~Special Galileon~~&~~$n^{\textrm{NLSM}}(k)$ [scalar]~~&~~$n^{\textrm{NLSM}}(k)$ [scalar]~~\\ 
\bottomrule
\end{tabular}
\end{center}
\end{table*}

 The striking dual-role between the kinematic and charge graph weights invites potentially fundamental as of yet unanswered questions.  What does it mean that we can treat gravitons as gluons whose charge is the kinematics of gluons?  What does it mean to think of Born-Infeld photons as gluons whose charge is the kinematics of pions? While the conceptual implications of color-dual double-copy structure has yet to be fully realized, the technical impact of the resulting algebraic constraints has however already been far reaching.  It allows perturbative calculations in quantum-gravity to achieve loop orders previously believed to be entirely out of reach by reframing them as predictions of much-more tractable quantum Yang-Mills calculations  \cite{BCJLoop}\cite{GeneralizedDoubleCopy,Bern:2018jmv}. These ideas have since been extended far beyond their original domain of on-shell scattering amplitudes to include form factors\cite{FourLoopFormFactor, FiveLoopFormFactor}, analysis of symmetries in gravity theories~\cite{Borsten2013bp, Borsten2013bp,
Anastasiou2014qba, Nagy:2014jza, Anastasiou2015vba, Cardoso2016amd,
Cardoso2016ngt, Anastasiou:2018rdx}, classical scalar, gauge, and Gravity solutions \cite{NeillRothstein,Monteiro2014cda,Luna2015paa,Ridgway2015fdl,Luna2016due,Luna2016hge,White2016jzc,Luna:2018dpt,Luna2017dtq,Goldberger2017frp,Goldberger2016iau,Goldberger2017vcg,CarrilloGonzalez2017iyj,Gurses:2018ckx}, and very recent implications for GR wave calculations.  Indeed the highest order post-Minkowski corrections to classical binary black-hole dynamics to date (3PM) has been carried out by leveraging these fundamental structures to fix coefficients in an effective action~\cite{3PM,Bern:2019crd}.

It is not clear what symmetry is responsible for this structure, nor the algebra the various kinematic graphical weights are charged under, not to mention the physical implications for the building blocks of fundamental theories. We therefore initiate an exploration of this last question, framing it in the context of on-shell constraints: {\em What (local) objects can satisfy the color kinematic duality?}

We discover that in certain cases this condition is more constraining than previously thought. Applying it to the simplest kinematic example, we find that it uniquely fixes the NLSM amplitudes and some of its higher derivative corrections.

It is also noteworthy that both open and closed string theory amplitudes at tree-level manifest field-theoretic adjoint color-dual double-copy structure.  The fact that double-copy seems compulsory to the effective building blocks of the only known ultra-violet completions of (higher-dimensional) Yang-Mills and gravity is incredibly tantalizing and suggests a compatibility with UV completion.  This further motivates the UV exploration that we will carry out.

\bigskip
The paper is organized as follows. In section 2 we briefly introduce the theories under consideration: bi-adjoint scalar, NLSM, DBI-VA, Gal, and sGal. In Section 3 we describe the on-shell constraints which will be used: locality, unitarity, soft limits, and the UV probes: the two particle BCFW shift, for both scalars and vectors, and the single hard limit, which can be understood as an off-shell BCFW shift. In Section 4 we expand on the amplitude relations, and show that NLSM is fixed by amplitude relations, locality, and mass dimension.  In Section 5 we present several uniqueness conditions that follow from imposing UV constraints, and mass dimension considerations:
\begin{itemize}
\item bi-adjoint doubly-ordered amplitudes from locality 
\item NLSM and BI from locality and unitarity
\item sGal from locality, unitarity, and a bound on the UV scaling of the Gal vertex
\end{itemize}
We  present evidence that uniqueness can still follow even after dropping the unitarity constraint. 
We summarize and discuss the outlook for future results in section 6.

\def\Tadj{\f}
\def\tree{{\rm tree}}
\def\f{ f}

\section{Field Theories}
\subsection{Conventions}
In this article we elide both phases and coupling constants to minimize unnecessary clutter.  An often used convention is to delegate coupling constants to full amplitudes (not ordered amplitudes), and set graph weights in accordance with the following double-copy prescription for full adjoint double-copy amplitudes as:
\begin{equation}
i^{L+1}{\cal A}^{\textrm{Dbl.cpy}(L)}_{n} =  g^{n-2+2L}_{m\otimes \tilde{m}}  \sum_{i}\, 
 \int \frac{d^{LD}\ell}{(2\pi)^{LD}} \frac{1}{S_i} \frac{m_{i} \tilde{m}_i}{D_i} \, ,
\end{equation}
where the sum is over all cubic graphs of loop order $L$ and multiplicity $n$, both $m$ and $\tilde{m}$ obey adjoint color-relations, namely Jacobi and antisymmetry, the $D_i$ are the massless propagators, $S_i$ are any symmetry factors of graph $i$, and $g_{m\otimes \tilde{m}}$ is the coupling constant for the theory.  The double-copy construction  then specifies the necessary scaling of coupling constants, e.g.~$g_{\textrm {YM}} = g$ and $g_{\textrm {GR}} = \kappa/2$, with associated phases and factors of $\sqrt{2}$ in color-factors and traces.  A thorough treatment of such a convention is given in ref.~\cite{bcjReview}.  

\subsection{Bi-adjoint $\phi^3$}
The bi-adjoint scalar theory (see e.g.~\cite{Bern1999bx,Du2011js,OConnellAlgebras,Cachazo:2013iea, Chiodaroli2014xia, Chiodaroli2015rdg, Low:2017mlh,White2016jzc} and references therein) is the simplest  theory within the web of amplitudes we are discussing, and provides the clearest formal access to fundamental structure at the heart of all the predictions within this web. It is a theory of scalars which carry two colors, and have a simple $\phi^3$ interaction.
\begin{equation}
\mathcal{L}=\frac{1}{2} \partial^{\mu} \Phi^{a a^{\prime}} \partial_{\mu} \Phi^{a a^{\prime}}+\frac{1}{3} f^{a b c} \tilde{f}^{a^{\prime} b^{\prime} c^{\prime}} \Phi^{a a^{\prime}} \Phi^{b b^{\prime}} \Phi^{c c^{\prime}} \, .
\end{equation}
 The full amplitudes can be decomposed into doubly-ordered partial amplitudes,
\begin{align}
\mathcal{A}&=\sum_{\sigma,\rho\in S_{n-1}} \!\!\!\!\!\textrm{Tr}(T^{a_1}T^{a_{\sigma(2)}}\ldots T^{a_{\sigma(n)}}) \times A^{\textrm{bi-Adj}}_n(1,\sigma |1,\rho) \times
\textrm{Tr}(\tilde T^{b_1}\tilde T^{b_{\rho(2)}}\ldots \tilde T^{b_{\rho(n)}})\\
&=\sum_{\sigma,\rho\in S_{n-2}}  \!\!\!\!\! c(1|\sigma|n)   \,A^{\textrm{bi-Adj}}_n(1,\sigma,n|1,\rho,n)  \,\tilde{c}(1|\rho|n) \, .
\end{align}
These doubly-ordered partial amplitudes $A_n(\sigma|\rho)$ are not unique to any theory, rather as we will see they encode the compatibility of cubic propagator structure with Jacobi satisfying graphical weights.  Indeed the $c(\sigma)$, $\tilde{c}(\rho)$ functions are simply the color-weights given to the half-ladder graphs of permutation $\sigma$ and $\rho$ by dressing each vertex with the  Lie-Algebra structure constants $f^{abc}$, $\tilde{f}^{a'b'c'}$.   Why is this sufficient?  Jacobi relations express color-charges for the set of all distinct $(2m-5)!!$ $n$-point graphs in terms of a basis set of $(n-2)!$ half-ladder (also called multiperipheral) graphs with the two farthest leg-labels fixed~\cite{DixonMaltoni}. 

Introducing a propagator containing matrix $P$ defined to be the  $(2n-5)!! \times (2n-5)!!$ diagonal matrix defined as $P_{ij}=\delta_{i,j} \frac{1}{D_i}$ where each non-vanishing element is the product of propagators for a particular graph,  the full amplitude can be written explicitly in terms of all graphs as 
\begin{equation}
\mathcal{A} = \sum_{i=1}^{(2n-5)!!} \frac{ c_i  \tilde{c}_i}{D_i} = c_{\textrm{all}} \cdot P \cdot \tilde{c}_{\textrm{all}} \, .
\end{equation}
The Jacobi solution matrix  $J$ matrix is an $(2m-5)!! \times (m-2)!$ matrix encoding how every graph's color factor is expressed via Jacobi relations terms of a basis of $(m-2)!$ master graph color factors: $(c_{\textrm{all}}) = J \cdot (c_{\textrm{masters}})$.  This makes it clear that the doubly-ordered bi-adjoint amplitude can be written:
\begin{equation}
A^{\textrm{bi-Adj}}= J^T \cdot P \cdot J \,,
\end{equation}
where the matrix indices of the doubly-ordered amplitude $A^{\textrm{bi-Adj}}$ index into a lexicographic ordering of the permutations $\sigma$ and $\rho$ -- specifying what master graphs define $J$.    

It is worth  considering an example at 4-points.   With the following definitions for the three cubic color weights:
\begin{align}
c_s &= c(1|23|4)=f^{a_1a_2 b}f^{b a_3 a_4}\,, \\
c_t &= f^{a_4 a_1 b}f^{b a_2 a_3}\,, \\
c_u &= c(1|32|4)= f^{a_1 a_3 b}f^{ b a_2  a_4 }\,,
\end{align}
and $\tilde{c}_i=c_i|_{f\to\tilde f}$, satisfying Jacobi: $c_t=c_s-c_u $ and  $\tilde{c}_t=\tilde{c}_s-\tilde{c}_u$.  The kinematic propagators are specified by Mandelstam variables, in an all out-going convention:
\begin{align}
s &= (k_1+k_2)^2=(k_3+k_4)^2\,,  \\
 t &= (k_1+k_4)^2=(k_2+k_3)^2\,,  \\
 u&= (k_1+k_3)^2=(k_2+k_4)^2 ,
\end{align}
satisfying $s+t+u=0$.
So our full amplitude is given:
\begin{align}
\cal{A} &= \frac{c_s \tilde{c}_s }{s}+\frac{c_t \tilde{c}_t }{t} +\frac{c_u \tilde{c}_u }{u}\\
&= \begin{bmatrix}
c_s & c_t & c_u
\end{bmatrix}  \cdot
\begin{bmatrix}
\frac{1}{s} & 0 & 0\\
0 & \frac{1}{t} & 0\\
0 & 0 & \frac{1}{u} 
\end{bmatrix}  \cdot
\begin{bmatrix}
\tilde  c_s\\
\tilde{c}_t\\
\tilde{c}_u
\end{bmatrix}\\
&= \begin{bmatrix}
c_s & c_u
\end{bmatrix}  \cdot
 \begin{bmatrix}
1 &1 & 0\\
0 & -1 & 1\\
\end{bmatrix} \cdot
\begin{bmatrix}
\frac{1}{s} & 0 & 0\\
0 & \frac{1}{t} & 0\\
0 & 0 & \frac{1}{u} 
\end{bmatrix} \cdot
\begin{bmatrix}
1&0\\
1&-1\\
0&1
\end{bmatrix} \cdot
\begin{bmatrix}
\tilde{c}_s\\
\tilde{c}_u
\end{bmatrix}\\
&= \begin{bmatrix}
c_s & c_u
\end{bmatrix}  \cdot
 \
\begin{bmatrix}
\frac{1}{s} +\frac{1}{t} & - \frac{1}{t} \\
-\frac{1}{t} & \frac{1}{t} +\frac{1}{u} 
\end{bmatrix}  \cdot
\begin{bmatrix}
\tilde{c}_s\\
\tilde{c}_u
\end{bmatrix}\\
&= c^{\rm T}_{\textrm{masters}} \cdot A^{\rm bi-Adj} \cdot  \tilde{ c}_{\textrm{masters}}  \, .
\end{align}

We will follow the convention in the literature and refer to these doubly-ordered quantities as {\em bi-adjoint} amplitudes, but wish to emphasize that they are far more universal than bi-adjoint  scalar $\phi^3$.  Only when dressing with two color-weights does the doubly-ordered bi-adjoint amplitude build the bi-adjoint scalar amplitude.  Every one of the amplitudes in the adjoint web of theories can be expressed by replacing these master graph color-weights with color-dual kinematic weights $m$ or $\tilde{m}$ as per Table~\ref{JacFactors},
\begin{equation}
\label{adjointDblCopy}
\mathcal{A}^{\textrm{Dbl.Copy}}\!\!=\!\!\!\!\!\!\!\sum_{\sigma,\rho\in S_{n-2}}  \!\!\!\!\!m(1|\sigma|n)   \,A^{\textrm{bi-Adj}}(1,\sigma,n|1,\rho,n)  \,\tilde{m}(1|\rho|n)\,.
\end{equation}
One can wonder if the doubly-partial bi-adjoint amplitudes have a simple closed form expression, and indeed they do, given as the inverse of the KLT matrix \cite{Cachazo:2013iea,Mizera:2016jhj} (c.f.~Eqn.~\ref{KLTmatrix}).  This is simplest to see in the case of an $(n-3)!$ basis, where the KLT matrix has a trivial inversion.  There is however no  real barrier to the $(n-2)!$ symmetric KLT matrix, but there is a subtlety as its inversion requires regulation on-shell.  A procedure of inverting off-shell,  canceling the on-shell singularity with $(k_n)^2$, and only than taking the on-shell limit $(k_n)^2\to0$ is similar to that discussed in e.g.~ref.~\cite{BjerrumBohr2010ta} as its intimately related to finding local color-dual kinematic weights in terms of ordered amplitudes using KLT.  As the double adjoint-striation collects trivalent graph propagators in terms of their dual-Jacobi master dressings, these objects are at the heart of adjoint-double copy.

In the simplest case when $\sigma=\rho$, the partial amplitude is simply the sum over the propagators of cubic graphs consistent with that color-order:
\begin{align}
A^{\textrm{bi-Adj}}_4(1,2,3,4|1,2,3,4)&=\frac{1}{s_{12}}+\frac{1}{s_{14}} \, ,\\
A^{\textrm{bi-Adj}}_5(1,2,3,4,5|1,2,3,4,5)&=\frac{1}{s_{12}s_{34}}+\frac{1}{s_{23}s_{45}}
+\frac{1}{s_{34}s_{51}}
+\frac{1}{s_{45}s_{12}}+\frac{1}{s_{51}s_{23}} \, .
\end{align}
In other cases, it is given by the set of propagators common to both orderings.

\subsection{NLSM}
The non-linear sigma model (NLSM) \cite{Cronin:1967jq}\cite{Weinberg:1966fm}\cite{Weinberg:1968de} is a pionic theory of Nambu-Goldstone bosons, which arises from spontaneously breaking a Lie group $G\times G\rightarrow G$.   It can be described by a Lagrangian in the Cayley
parameterization~\cite{macfarlane1968,Kampf:2013vha,Carrasco2016ldy,LowSoft},
\begin{equation} 
\label{NLSMaction}
{\cal L}_{{\rm NLSM}} = {1\over 2} {\rm Tr} \bigg\{ \partial_\mu \varphi \, 
{1\over 1- \lambda  \varphi^2} \, 
  \partial^\mu \varphi \, {1\over 1 - \lambda\varphi^2} \bigg\} \, ,
\end{equation}
where $\varphi$ is a Lie-algebra valued Goldstone-boson scalar field in the adjoint representation. 

We will focus on the $SU(N)$ NLSM amplitudes $\mathcal{A}_n$, which can be decomposed into flavor-ordered ``partial amplitudes'' $A_n$:
\begin{align}
\label{trace}
\mathcal{A}_n=\!\!\!\!\!\sum_{\sigma\in S_{n-1}}\!\!\!\!\!\textrm{Tr}(T^{a_1}T^{a_{\tau(2)}}\ldots T^{a_{\sigma(n)}})A(1,\tau(2,\ldots,n)) \, .
\end{align}

The four and six point ordered amplitudes read:
\begin{align}
A_4^{\textrm{NLSM}}(1,2,3,4)&=s_{13} \, ,\\
A_6^{\textrm{NLSM}}(1,2,3,4,5,6)&=\frac{s_{13}s_{46}}{s_{123}}-s_{13}+\textrm(cyclic) \, .
\end{align}
The Lagrangian enjoys a shift symmetry, equivalent to the Adler zero condition (see  \cite{Cheung:2016drk,Low:2017mlh}), which in turn is sufficient to fully determine the on-shell amplitudes.  

\subsection{DBI-VA}
We will also consider the DBI-VA model. This is a nonlinear extension of Maxwell theory, which in D dimensions is given by the following Lagrangian
\begin{align}
\label{BI}
\mathcal{L}_{\textrm{BI}}=\sqrt{(-1)^{D-1}\textrm{det}(\eta_{\mu\nu}+F_{\mu\nu})} \, .
\end{align}
Its supersymmetric extension has also been considered \cite{Bagger:1996wp,Bergshoeff:2013pia}.

The scalar part, known as DBI, is fixed by a stronger Adler zero condition $\mathcal{O}(\tau^2)$, which similarly follows from a more general shift symmetry.
Intriguingly, for the vector part, known as BI, in \cite{CheungSoft} it was shown that starting from a general Lagrangian
\begin{align}
\label{BIform}
\mathcal{L}_{\textrm{BI}}=F^2+g_4F^4+g_6F^6+\ldots \, ,
\end{align}
and demanding an improved low energy behavior, the coefficients $g_i$ can be fixed to match the expansion of (\ref{BI}). 4D amplitudes in the vector, fermion, and respectively scalar sector are given by:
\begin{align}
A_4^{\textrm{BI}}(\gamma_1^-,\gamma_2^-,\gamma_3^+,\gamma_4^+)&=\langle 12\rangle^2[34]^2 \, ,\\
A_6^{\textrm{BI}}(\gamma_1^-,\gamma_2^-,\gamma_3^-,\gamma_4^+,\gamma_5^+,\gamma_6^+)&=\frac{\langle 12\rangle^2 [56]^2 \langle 3|1,2|4]^2}{s_{124} }
+\textrm{perms.} \, ,\\
A_4^{\textrm{VA}}(\psi_1,\psi_2,\bar{\psi}_3,\bar{\psi}_4)&=\langle 12\rangle[34]s_{12} \, ,\\
 A_6^{\textrm{VA}}(\psi_1,\psi_2,\psi_3,\bar{\psi}_4,\bar{\psi}_5,\bar{\psi}_6)&=\frac{\langle 12\rangle [56] \langle 3|1,2|4]s_{12}s_{56}}{s_{124}}
+\textrm{perms.} \, ,\\
A_4^{\textrm{DBI}}(\phi_1,\phi_2,\phi_3,\phi_4)&=s_{12}^2+s_{23}^2+s_{13}^2 \, ,\\
A_6^{\textrm{DBI}}(\phi_1,\phi_2,\phi_3,\phi_4,\phi_5,\phi_6)&=-s_{12}s_{34}s_{56}\nonumber\\
+(s_{12}^2+s_{23}^2+s_{13}^2)&(s_{45}^2 +s_{56}^2+s_{46}^2)\frac{1}{s_{123}}
+\textrm{perms} \, .
\end{align}
The general dimension BI amplitudes can be laborious to write out even at four points, but they can be given by:
\begin{equation}
A_4^{\textrm{BI}}=s t A_4^{\textrm{YM}}= \big[ 4{\rm Tr}(F_1F_2F_3F_4) -
 {\rm Tr}(F_1 F_2)  {\rm Tr}(F_3 F_4) + {\rm cyclic}(1,2,3)\big]\,.
\end{equation}
The  traces are over Lorentz  indices of linearized momentum-space field strengths:
\begin{equation}
F^{\mu\nu}_i \equiv p^{\mu}_i \epsilon^{\nu}_i- \epsilon^{\mu}_i p^{\nu}_i\,.
\end{equation}

More generally, the full supersymmmetric DBI-VA amplitudes can be obtained via the double-copy procedure as:
\begin{align}
A^{\textrm{DBIVA}}=A^{\textrm{SYM}}\otimes A^{\textrm{NLSM}} \, ,
\end{align}
more precisely given in terms of partial amplitudes by eq. (\ref{KLT}) or in terms of color-dual dressed cubic graphs as per eqn.~\ref{adjointDblCopy}.

\subsection{Gal and sGal}
The Galileon is a theory of scalars which originally appeared in the context of gravity models  \cite{Dvali:2000hr,Nicolis:2008in,deRham:2010kj}, also discussed in \cite{Kampf:2014rka}. It is given by a Lagrangian of the form
\begin{align}
\label{gall}
\mathcal{L}_{\textrm{Gal}}=-\frac{1}{2}(\partial \phi)^2+(\partial \phi)^2\sum_{n=4}^{\infty} c_n \textrm{det}_n \, ,
\end{align}
where $\textrm{det}_n=n!\partial^{[\mu_1}\partial_{\mu_1}\phi \ldots \partial^{\mu_n]}\partial_{\mu_n}\phi$.   The contact terms, or Galileon vertices, are neatly given by:
\begin{align}
\label{defgal}
V_n=\textrm{Det}(M^{a}) \,,
\end{align}
where $M^{a}$ is the $(n-1)\times(n-1)$ matrix obtained by removing any row $a$ and column $a$ from the matrix $M_{ij}=p_i.p_j$, $i,j=\overline{1,n}$. Although not obvious, permutation invariance follows from momentum conservation.  Its scattering amplitudes are given by:
\begin{align}
A_4^{\textrm{Gal}}&=c_4 V_4 \, ,\\
A_5^{\textrm{Gal}}&=c_5 V_5 \, ,\\
A_6^{\textrm{Gal}}&=(c_4)^2\left(\frac{V_4(1,2,3,p)\times V_4(-p,4,5,6)}{s_{123}}+\textrm{perms.}\right)
+c_6V_6 \, .
\end{align}

Finally, the special Galileon is a particular linear combination of the Galileon operators, which satisfies an even stronger Galileon symmetry \cite{Hinterbichler:2015pqa,Novotny:2016jkh}, as well as a stronger $\mathcal{O}(\tau^3)$ Adler zero condition. Unlike the general Galileon, the special Galileon amplitudes can also be obtained via the adjoint double-copy procedure as:
\begin{align}
A^{\textrm{sGal}}=A^{\textrm{NLSM}}\otimes A^{\textrm{NLSM}}\,.
\end{align}

\section{Amplitude Constraints}
\subsection{Locality}
Locality fixes\footnote{We note that there is a certain ambiguity, present in the literature, regarding whether it is locality or unitarity that requires simple poles. Here we will use the term {\em locality} to denote a constraint on the properties of denominators, i.e. the presence of only simple poles, and the term {\em unitarity} to constrain the factorization properties of amplitudes on these simple poles.} the pole structure of the functions considered, and will be assumed throughout the paper. A local ansatz may be written as:
\begin{align}
B_n^{\textrm{local}}=\sum_i \frac{N_i}{D_i} \, ,
\end{align}
where, depending on the theory considered, the sum runs over all over cubic or quartic tree diagram topologies $i$ with corresponding massless scalar propagators $D_i$. The $N_i$ are polynomials of momenta (and polarization vectors for BI), with unfixed coefficients, with their mass dimension fixed in terms of the net mass dimension of the amplitude. For NLSM, BI, and Gal, the quartic structure implies that each diagram in the ansatz will have exactly $n/2-2$ poles, thus fixing the mass dimension of the numerators to $[n-2]$, $[2n-4]$, and $[3n-6]$ respectively. Terms with fewer (or zero) poles, such as contact terms, are (non-uniquely) included in the numerators. Since at this stage we are not yet assuming unitarity, these $N$ do not have any initial factorization properties.

We  will assume for full generality and to maximize potential independence of kinematic invariants, that unless otherwise specified the spacetime dimension can be taken arbitrarily large, at least $D_{\textrm{ST}}>n$ for any $n$-point amplitude in consideration. 
\subsection{Unitarity}
Unitarity further imposes that, on each pole $P^2$, $B_n^\textrm{local}$ factorizes into two lower point amplitudes:
\begin{align}
\label{unitarity}
\lim_{P^2\rightarrow 0}B_n=\frac{A_L\times A_R}{P^2} \, ,
\end{align} This implies that with unitarity the only unfixed piece of the ansatz is a potential contact term:
\begin{align}
B_n^{\textrm{unitary}}=\textrm{[factorizing piece]}+C_n(p^m) \, ,
\end{align}
where the first part is fully determined by eq. (\ref{unitarity}), and $C_n$ is now a polynomial of mass dimension $[m]$, with unfixed coefficients. 

We find demanding unitarity an extremely sharp constraint that will allow us to prove many uniqueness claims. In most cases we will find additional evidence that unitarity surprisingly emerges as a consequence of locality and other properties. 

\subsection{Soft Limits}
Soft limits will be central to our arguments for uniqueness. Given their universality and usefulness, soft theorems have been under intensive recent study (see e.g. \cite{He:2014bga,He:2016vfi,Guerrieri:2017ujb,Bern:2014vva,Bern:2014oka,Huang2015sla,Chen:2014xoa,Low:2015ogb,Elvang:2016qvq,DiVecchia:2015jaq,DiVecchia:2015oba,Schwab:2014xua,RodinaSoft}\cite{Strominger:2017zoo} and references therein). We note however that we will not be assuming or imposing any of the soft theorems, but only using soft limits as formal Taylor expansions. In some cases, the soft theorems will in fact arise from UV constraints.

There are two closely related types of soft limits that we will use. First is the Adler zero, which involves taking one particle soft by rescaling one momenta $p\rightarrow \hat{p}=\tau p$, and taking $\tau\rightarrow 0$. In this limit, several special EFT's scale as:
\begin{align}
A\rightarrow \mathcal{O}(\tau^\sigma) \, ,
\end{align}
where $\sigma=1$ NLSM, $\sigma=2$ for DBI and Galileon, and $\sigma=3$ for sGal \cite{Cheung:2016drk}. These particular values for $\sigma$ are interesting because they are below what simple mass dimension counting would imply. Take for instance the NLSM at 6 points:
\begin{align}
A_6=\frac{s_{13}s_{46}}{s_{123}}-s_{46}+(\textrm{cyclic}) \, .
\end{align}
While each term separately scales as $\mathcal{O}(\tau^0)$ under a $p_2\rightarrow 0$ limit, their sum has an improved $\mathcal{O}(\tau^1)$ scaling. Such cancellations become highly non-trivial at higher points and for other theories like DBI or sGAL, and are in fact so powerful they fully constrain the theories \cite{Cheung:2014dqa,Arkani-Hamed:2016rak,RodinaGaugeInv}.

The other type of soft behavior relevant for EFT's is the double soft expansion: \cite{ArkaniHamed2008gz,Cachazo:2015ksa}:
\begin{align}
A_{n+2}\rightarrow  \tau^\sigma (S_0+\tau S_1+\ldots)A_n  \, .
\end{align}
In this case the non-trivial aspect is the factorization between the ``soft factors'' $S_i$ and the lower point amplitude $A_n$. Like the Adler zero, this expression places very stringent constraints on the amplitudes, in fact again sufficiently strong to fully constrain them \cite{RodinaSoft}. For EFT's, this later claim relies on the following fact: there are no objects with enhanced double soft limits, except for the Galileon vertices. 

These ``uniqueness'' results can be turned into very powerful tools, as they imply amplitudes are fully determined by just the first few orders in a soft expansion. Not only does this greatly simplify checks, in many cases it facilitates proofs through inductive arguments. Since we will use these results throughout the article, we can can rephrase them more succinctly and practically:

\begin{itemize}
\item There are only four local objects which have enhanced single soft limit: NLSM, DBI, Galileon vertex, sGal. 
\item There is a unique local object which has enhanced double soft limit: the Galileon vertex. 
\item Anything else has a scaling dictated purely by mass dimension and singularity structure.
\end{itemize}
We should mention that these facts have not been proven rigorously for the Galileon or the special Galileon, but such proofs likely follow from arguments of the type given in \cite{RodinaGaugeInv}. For completeness, we will prove one particular case which shows up when discussing BI:
\begin{itemize} 
\item {\em There is no polynomial of mass dimension $[n]$ with double soft scaling $\mathcal{O}(\tau^3)$}
\end{itemize}

\subsection{2S: Two-particle-shift Scaling}
The BCFW shift \cite{BCFW} was originally introduced in four dimensions to enable a powerful on-shell recursion. Briefly, the recursion relies on using Cauchy's residue theorem to rebuild amplitudes from lower point information via unitarity. In $D$-dimensions, this is achieved via a scalar shift:
\begin{align}
\label{bcfw}
p_i&\rightarrow p_i+z q \, ,\\
p_j&\rightarrow p_j-z q \, ,
\end{align}
subjected to $p_i.q=p_j.q=q^2=0$, needed to preserve the on-shell conditions, or a vector shift \cite{Rodina:2016mbk}:
\begin{align}
\label{vbcfw}
\epsilon_i&\rightarrow \hat{\epsilon}_i \, ,\\
\epsilon_j&\rightarrow \hat{\epsilon}_j+z p_i\frac{\hat{\epsilon}_i.\epsilon_j}{p_i.p_j} \nonumber \, ,\\
p_i&\rightarrow p_i+z \hat{\epsilon}_i  \nonumber \, ,\\
p_j&\rightarrow p_j-z \hat{\epsilon}_i  \nonumber \, ,
\end{align}
where $\hat{\epsilon}_i=\epsilon_i-p_i\frac{\epsilon_i.p_j}{p_i.p_j}$. In both cases we will refer to shifts as $[i,j\rangle$.

If the amplitude vanishes for large $z$, it can be rebuilt purely from its residues in an extremely efficient manner:
\begin{align}
A_n=\sum_k \frac{A_L(z_k)A_R(z_k)}{P^2} \, ,
\end{align}
where the sum runs over all channels where $P^2(z_k)=0$. Even if the amplitude does not vanish at large $z$, the recursion may be generalized to multi-line shifts, and complemented by other properties, like the Adler zero for EFT's \cite{Luo:2015tat,Cheung:2016drk,Kampf:2013vha,Elvang:2018dco,Low:2019ynd}, see also \cite{Du2011js,He:2018svj} for bi-adjoint scalar amplitudes. In any case, the scaling is crucial, but difficult to compute. This is because very complicated cancellations occur such that the actual scaling is well below the naive expectation from power counting. The fact that these cancellations occur at all seems almost miraculous, and is a fact completely hidden from the Lagrangian perspective. The most well known scalings are for YM and GR, which behave as \cite{ArkaniHamed:2008yf,Benincasa:2007qj,Schuster:2008nh}
\begin{numcases}
{A^{\textrm{YM}}\sim}
\label{ymz}
  \mathcal{O}(z^{-1}), &\textrm{ for adjacent $i,j$}\\
  \mathcal{O}(z^{-2}), & \textrm{ for non-adjacent $i,j$}
\end{numcases}
\begin{align}
A^{\textrm{GR}}&\sim \mathcal{O}(z^{-2}) \, ,
\end{align}
making them perfect candidates for the BCFW recursion. 
In this article we find that EFT's also have an enhanced scaling at large $z$:
\begin{numcases}
{A^{\textrm{NLSM}}\sim}
\label{nlsm2s}
  \mathcal{O}(z^1), &\textrm{ for adjacent $i,j$}\\
  \mathcal{O}(z^0), & \textrm{ for non-adjacent $i,j$}
\end{numcases}
\begin{align}
A^{\textrm{BI}}&\sim  \mathcal{O}(z^{0}) \, ,\\
\{A^{\textrm{DBI}}, A^{\textrm{Gal}}, A^{\textrm{sGal}}\}&\sim \mathcal{O}(z^{2}) \, .
\end{align}
We have checked each of these scalings explicitly through various low-multiplicities.  It turns out that because of their double-copy structure, to know their 2S scaling at all multiplicity, one  only needs to know the scaling of their building blocks.  Namely, knowing the scaling of NLSM and YM amplitudes are sufficient for all of the above amplitudes.  All multiplicity Yang-Mills scaling is constrained by generalizations of  action arguments found in Ref.~\cite{ArkaniHamed:2008yf}, and while we expect that a similar all orders argument exists for the NLSM, we do not pursue it here.  Instead we simply verified NLSM BCFW scaling explicitly through 10 points.  

The point of this article is however not to apply the recursion and construct amplitudes directly via unitarity, but instead show that particular amplitudes can be completely defined by demanding enhanced UV behavior, and that unitarity follows as a consequence. This result is particularly surprising for EFT's, for the following reason. In gauge and gravity theories, we have already seen that gauge invariance completely fixes the form of the amplitudes (both with unitarity \cite{Boels:2016xhc}, and without \cite{Arkani-Hamed:2016rak,RodinaGaugeInv}). Given that the vector shift (\ref{vbcfw}) seems to incorporate a gauge transformation, it is not difficult to believe the enhanced scaling implies gauge invariance, and hence fixes the amplitudes.
For the EFT's under consideration however, it is the Adler zero that fixes contact terms. Yet how the Adler zero might be encoded in the shift (\ref{bcfw}) is even more mysterious. Not to mention that BI (as the off-spring of YM and NLSM - prime representatives of gauge invariance and Adler zero, respectively) is not fixed even by both, at least not directly. Via dimensional reduction, it was argued in \cite{CheungSoft} that BI may be fixed by the combination of Adler zero and gauge invariance. That the constraints following from such a complicated procedure can simply be turned into UV conditions is nevertheless quite surprising.

\subsection{SHS: Single-Hard Scaling}\label{shss}

The existence of on-shell recursion~\cite{BCFW,CSW} for special constructible theories suggests there is value in considering the constraining information via on-shell quantities such as the BCFW shift we consider above.  Given the utility of venerable Berends-Giele off-shell recursion approaches and related perturbiner methods (e.g. refs.~\cite{Berends:1987me,Lee:2015upy,Mafra:2016ltu,Mizera:2018jbh}) for scattering, it is perhaps a natural question to ask whether an off-shell constraint may provide sufficient information for a bootstrap constructibility program. We now introduce one such off-shell constraint, in the form of a single off-shell ``hard'' limit. We will see that when combined with unitarity/factorization, at a specified mass-dimension, the described single-hard scaling can completely constrain the predictions of many theories.  More tantalizing through explicit calculation in these theories at various accessible multiplicities, we find evidence to support conjectures that the single hard-scaling alone can be sufficient to entirely constrain these amplitudes, in some sense allowing unitarity to emerge from such considerations.  

Let us consider an amplitude:
$A(\ldots p \ldots)$ where we  want to take a single leg hard via:
$\hat{A_n}(\ldots \hat{p} \ldots)$ via a rescaling $p\to\hat{p}=z p$.  We have to be careful because the momentum conserving delta function is in a sense trivialized in this limit
\begin{align}
\delta(z p+\sum_i^{(n-1)} p_i)\rightarrow \delta(z p) \,.
\end{align}
and the remaining momenta are poorly constrained, in contrast to the case of taking $z p$ soft, when momentum conservation can be dealt with consistently \cite{Broedel:2014fsa}.  As one can use conservation of momenta to obscure the scaling of $p$, to unambiguously define a scaling, in a similar manner as when defining soft-limits, we insist on using a $p$-favoring basis of momentum invariants that makes the $p$ dependence of $A_n$ explicit.  Doing so requires only specifying a leg $i$ to always eliminate in favor of $p$, as well as a distinct momentum invariant $p_j . p_k$ where $p_j\ne p_k \ne p_i \ne p$ to also be eliminated from the basis of invariants.  As such one  can label any a set of basis of momentum invariants that satisfy conservation of momentum and maximally favor the appearance of $p$ by a triplet: $\delta(i,[j,k])$  with the following defining properties.
\begin{itemize}
 \item $\delta(i,[j,k])$ is any basis of momentum invariants that  explicitly removes any reference to $p_i$ and $p_j . p_k$ in its  basis elements.
 \item This can be accomplished by solving  the set of equations generated by considering both $ p_m^2 =0$ and $(\sum_{l=1}^n p_m . p_l )=0$ for every $m$, eliminating  $p_j.p_k$ and all $p_i . p_m$ in favor of other momentum invariants.
 \item Furthermore in the case of vector theories, enforcing $0=p_m . \epsilon_m$ for all $m$, and eliminating either $\epsilon_i . p_j$ or  $\epsilon_i . p_k$ in favor of $\epsilon_i . p$ via $ 0=\sum_{m=1}^n  p_m .\epsilon_i  $.
\end{itemize}
Once cast into an appropriate $p$-favoring basis of momentum invariants by applying $\delta(i,[j,k])$, the scaling of the hard-limit can unambiguously be extracted, but will depend on the relative positions of the hard particle and the three particles singled out by $\delta(i,[j,k])$. The hard particle $p$ and particle $p_i$ separate the ordered set $\sigma=\{1,\ldots,n\}$ into two parts, $L$ and $R$ (either possibly empty):
\begin{align}
A(p,L,p_i,R) \, .
\end{align}
Now, with respect to the separation $\sigma=(p,L,p_i,R)$, we define the set $\delta=(i,[j,k])$ as being: 
\begin{itemize}
\item compatible with $\sigma$, for $i$ not adjacent to $p$, and $\{j,k\}\in L$ or $\{j,k\}\in R$   
\item not compatible, otherwise
\end{itemize}
As a roadmap to the results presented in detail in \sect{UVBehaviorSection}, we will summarize here what we discover about the single-shift UV behavior of the theories under consideration. Through explicit calculation in accessible multiplicities, YM and NLSM present the following enhanced behavior for ordered amplitudes when taking $z p$ to be large:
\begin{numcases}{A_n(\sigma)\sim}
  \mathcal{O}(z^0),& \textrm{for compatible ordering}\\
  \mathcal{O}(z^1),& \textrm{otherwise}
\end{numcases}
The notation comes in handy for the bi-adjoint scalar, whose amplitudes are now:
\begin{align}
A(\sigma_1,\sigma_2)=A(p,A_1,p_i,B_1|p,A_2,p_i,B_2) \, .
\end{align}
With this notation, the bi-adjoint scalar scales as 
\begin{numcases}{A_n(\sigma_1|\sigma_2)\sim} \label{biadj}
  \mathcal{O}(z^{-3}),& \textrm{for $\delta$ compatible with both $\sigma_1$ and $\sigma_2$}\\
  \mathcal{O}(z^{-2}) \textrm{ (or better)},& \textrm{for $\delta$ compatible with either $\sigma_1$ or $\sigma_2$}\\
 \mathcal{O}(z^{-1}) \textrm{ (or better)},& \textrm{otherwise}
\end{numcases}
The bi-ordered scaling is not completely determined because we have not taken into account the relative ordering between $\sigma_1$ and $\sigma_2$. Amplitudes with more ``orthogonal'' relative orderings can contain very few terms, and in this case even bad choices for $\delta$ can have improved scaling, which requires no cross-term cancellations. However, the above minimum requirements will be sufficient for our purposes. 
Next, for the full color-dressed NLSM and YM amplitudes, as well as GR, DBI, sGal and the Gal vertex we find:
\begin{align}
\{{\cal{A}}^{\textrm{YM}},{ \cal{A}}^\textrm{NLSM}\}&\sim \mathcal{O}(z^{1}) \, ,  \\
\{A^{\textrm{GR}}, A^{\textrm{DBI}}, A^{\textrm{sGal}}\}&\sim \mathcal{O}(z^{3}) \, ,  \\
V^{\textrm{Gal}}&\sim  \mathcal{O}(z^{4}) \, ,
\end{align}
independent of the choice for $\delta(i,[j,k])$ due to permutation invariance. Although surprising, this independence will be easily understood through the double copy procedure. 
Furthermore, higher derivative corrections to any of these theories follow a similar pattern. Assuming $\kappa$ extra derivatives to NLSM, YM, GR, sGal, etc., we universally find:
\begin{align}
A_n^\kappa\sim z^{\kappa/2} A_n^{(\kappa=0)} \, .
\end{align}
It should be mentioned that, like for the two-particle shift, intricate and quite unexpected cancellations between different Feynman diagrams are required to enable these scalings. 

We have explicitly checked that each of these scalings hold through various low-multiplicities.  Because of their double-copy structure, as with the BCFW scaling, to know their single hard scaling at all multiplicity, one only needs to know the scaling of their building blocks, namely: NLSM and YM. In the single-hard scaling case we do not have all multiplicity Yang-Mills or NLSM single-hard scaling arguments available, and so explicitly verified Yang-Mills through $n=7$ and NLSM through $n=10$.  We believe it would be interesting to pursue all-multiplicity proofs of this novel single hard scaling.  Here we occupy ourselves with a different question: what additional constraints, beyond a particular SHS scaling (e.g. locality, unitarity, mass-dimension, etc), are required to uniquely specify the amplitudes of a given theory?  We address these questions on a case by case basis in \sect{UVBehaviorSection}. 

One peculiarity of this scaling is that for ordered amplitudes the (non)adjacency of $(n+1)$ and $i$ matters, similar to how it does for a BCFW shift.
Another obvious feature is that the single hard limit essentially looks like ``half'' of a BCFW shift. The exact difference is easy to quantify. The two-line particle shift (\ref{bcfw}) does not affect momentum conservation, so the overall scaling is independent of the triplet $\delta(i,[j,k])$. Choose $i=n$ and consider a $[1,n\rangle$ shift. Because $p_n$ does not appear explicitly, the shift actually reduces to a single deformation:
\begin{align}
p_1\rightarrow p_1+z q \, ,
\end{align}
subjected to $q.p_1=0$ and $q.(\sum_{i=1}^{n-1} p_i)=0$. It is now clear that to obtain the single hard limit we need only set $q=p_1$ and drop the on-shell condition $q.(\sum_{i=1}^{n-1} p_i)=0$. Conversely, a full momentum conserving two particle shift can simply be obtained from a single hard limit by imposing the extra on-shell condition $q.(\sum_{i=1}^{n-1} p_i)=0$. An immediate consequence of this fact is that the behavior under a two particle shift cannot be worse than under a single hard limit.

While apparently ill-defined because of momentum conservation issues, unlike the typical high energy limits involved in the Froissart bound \cite{Froissart:1961ux}, the single hard limit is nevertheless a valid and natural kinematic configuration to explore. It turns out to be a non-trivial property of many theories, and in fact a defining property of NLSM and sGal. It would be interesting nonetheless if the enhanced behavior has any implications for the Froissart bound itself.

\section{Amplitude Relations and the NLSM}
The trace basis (\ref{trace}) is not minimal, and can be further reduced two times. First, the  Kleiss-Kuijf (KK)
 amplitude relations~\cite{KleissKuijf}:
 \begin{equation}
 A_n(1, \alpha, n, \beta) = (-1)^{|\beta|} \sum_{\sigma \in \alpha \shuffle {\beta^T} } A_n(1, \sigma, n)\,, 
\end{equation}
where $\alpha$ and $\beta$ are lists of external labels, $\beta^T$ represents the reverse ordering of the list $\beta$, and $ \alpha \shuffle \beta^T$ are the permutations that shuffle the $\alpha$ and $\beta$, i.e. that separately maintain the relative order of the  elements belonging to each list but can interleave elements from both lists. These relate the $(n-1)!$ ordered amplitudes, $A(1,\tau)$, amplitudes to an $(n-2)!$ basis with two legs fixed: $A(1,\sigma, n)$.  Writing the adjoint generator matrices as $(\Tadj^a)_{bc} \equiv \f^{bac}$ one can
write any flavor factor as products of $\Tadj^{a_i}$'s.  This leads to the following expression for the full flavor-dressed amplitude~\cite{DixonMaltoni}:
\begin{equation}
{\cal A}_n^\tree = g^{n-2} \sum_{\sigma \in S_{n-2}}   
A_n^\tree \big(1,\sigma,n\big) 
c(1|\sigma | n) \,,
\end{equation}
where $c(1|\sigma |n)$ is the color-weight of the cubic half ladder graph with farthest legs $1$ and $n$ fixed and the intermediate legs labeled according to $\sigma$.
Second, because the NLSM in the adjoint obeys color-kinematics~\cite{Chen2013fya}, this implies a further reduction to a basis of $(n-3)!$ independent amplitudes. This necessarily manifests in the satisfaction of the simplest, or so-called {\em fundamental} BCJ, relations~\cite{BCJ,amplituderelationProof} which can be written as:
\begin{align}
\label{BCJ}
\sum_{i=2}^{n-1} k_{1i}A(2,\ldots ,i, 1,i+1,\ldots ,n)=0 \, ,
\end{align}
where $k_{1i}=\sum_{j=2}^{i}p_{1}\cdot p_{j}$.

A key consequence of the adjoint double-copy structure is that at tree-level color-dual kinematic graph weights can be given by specifying the kinematic numerators of $(n-2)!$ half-ladder master graphs, deriving all other weights by Jacobi. One representation for Jacobi-satisfying kinematic weights for the non-linear sigma model is simply:
\[
n^{NLSM}(1|\sigma|n) = S(\sigma | \sigma) \, ,
\]
where S is the celebrated $(n-2)!$ rank KLT matrix given below in eqn.~\ref{KLTmatrix}.  This form was conjectured in ref.~\cite{Carrasco2016ldy} and proven from string-theoretic considerations in \cite{Carrasco2016ygv}. This intimate relation between the KLT matrix and NLSM amplitude is a first hint of the deeper fact we prove here, that NLSM amplitudes are uniquely specified by the quartic structure and color-kinematics.

The NLSM further plays a central role in the other theories we will consider
 \begin{enumerate}
 \item supersymmetric DBI-VA :   NLSM $\otimes$ supersymmetric Yang-Mills
 \item Special Galileon:   NLSM $\otimes$ NLSM
 \end{enumerate}

Each of these theories has diverging high-energy behavior and so can be understood as effective field theories requiring some sort of completion in the UV. The NLSM itself finds a UV completion in abelian Z-theory~\cite{Carrasco2016ldy} and the (supersymmetric) DBI-VA has a UV completion in the abelian supersymmetric open-string.   While each of the double-copy factors above can admit higher-derivative corrections,  it is interesting to note that both Z-theory and the abelian open string only exploit higher-derivative corrections to their respective pion factors.  Indeed both UV completions receive higher derivative corrections to their pion factors in the same ratio, as the abelian open string at tree-level can be understood as a field theory double copy between abelian Z-theory amplitudes and supersymmetric-Yang-Mills. As we will see, the combination of locality, higher-derivatives, and amplitude relations in combination with unitarity can be highly constraining.

\subsection{Uniqueness from amplitude relations} 
Here we finally explore the space of local objects that can obey the color-kinematic duality, or equivalently, the amplitude relations. Imposing the amplitude relations rather than graph-level numerator constraints in this case is much more efficient. To compare, an ordered quartic amplitude scalar ansatz at 8 points has 18,540 terms, which are subject to just two constraints: cyclic invariance and the BCJ relations. On the other hand, a general cubic ansatz  for the 8-point half-ladder graph relevant to NLSM powercounting has 177,100 terms, which must satisfy anti-symmetry about each vertex and the Jacobi relations about each propagator, as well as vanishing of all cubic-residues of the resulting ordered amplitudes. 

\begin{claim}
Flavor-ordered (pionic) NLSM amplitudes are fixed uniquely by locality, [mass dimension] $=2$, cyclic invariance, and the BCJ relation:
\begin{align}
\label{fBCJ}
F_n=&\sum_{i=2}^{n-1} k_{1i}A_{i;n}(2,\ldots ,i, 1,\ldots ,n)=0 \, ,
\end{align}
where $k_{1m}=\sum_{i=2}^{m}p_{1}\cdot p_{i}$. 
\end{claim}
We do not need to assume that the $A_{i;n}$ are related to each other by relabeling, so this is in fact a stronger statement than the numerator level duality. To prove this claim we will make use of the Adler zero uniqueness. Taking the first particle soft as $p_1=\tau p_1$, with $\tau\rightarrow 0$, we require that $F_n$ vanishes order by order in $\tau$:
\begin{align}
F_n(\tau)\rightarrow \tau F^{(1)}+\tau^2 F^{(2)}+\ldots=0 \, ,
\end{align} 
which will involve the soft limit expansion of the amplitudes:
\begin{align}
A_n(\tau)\rightarrow \tau^0 A_n^{(0)}+\tau A_n^{(1)}+\ldots \, .
\end{align} 
Because of the quartic propagator structure, no pole in $A(n)$ is singular in this limit, hence at leading order the dependence on the soft momenta drops out:
\begin{align}
A_n(\tau)=\sum \frac{N(\tau p_1)}{D(\tau p_1)}=\sum \frac{N^{(0)}+\tau p_1 N^{(1)}+\ldots}{D^{(0)}+\tau P^{(1)}+\ldots}\rightarrow\frac{N(0)}{D(0)}+\ldots=A_n^{(0)}+\ldots \, .
\end{align}
Now we impose
\begin{align}
F^{(1)}=\sum_i k_{1i}A_{n;i}^{(0)}(1,\ldots,i,\ldots,n)=0 \, .
\end{align}
The $k_{1i}$ coefficients in front of each $A_i$ are independent (under $n$-point kinematics) and since the $A_i$'s themselves are independent of $p_1$, the above equation implies that each $A_n^{(0)}$ must vanish separately. But this is precisely the Adler zero condition for particle 1. Using cyclic invariance it means that each $A_i$'s must satisfy the Adler zero in all $n$ particles, and there is a unique  local object with this property: the NLSM amplitude.

\subsection{Higher derivative corrections from amplitude relations and unitarity}

The amplitude relations (\ref{fBCJ}) also put stringent constraints on higher derivative corrections to the NLSM~\cite{Elvang:2018dco}. While not sufficient just with locality, we find that also assuming unitarity can uniquely determine such amplitudes, to some finite mass dimension. To test this claim we can write an ansatz:
\begin{align}
\label{contactnlsm}
B_n^\textrm{NLSM+H.D.}=\textrm{[factorizing piece]}+C_n(p^\kappa) \, ,
\end{align}
where the factorizing part is fully determined by unitarity, while the contact term $C_n$ is a polynomial. If we consider theories coming from $\mathcal{O}(p^\kappa)$ operators, then this polynomial has mass dimension $[\kappa]$, with $\kappa=2$ corresponding to the usual NLSM amplitude. We have found that at 6 points, up to $\kappa=10$ the contact term cannot satisfy the BCJ relations on its own, and therefore unitarity plus amplitude relations fix the ansatz uniquely. For $\kappa=12$  there are two polynomial solutions to the BCJ relations.  

One has to go to $\kappa=14$ at four points to find the first analogous situation where two independent color-kinematic satisfying solutions occur, but it is instructive to explore.  Quite simply. the $s,t$  channel has the following two independent solutions:
\begin{equation}
A^{[14]}(s,t) = u \,  \left(  \alpha ( s \,t\, u )^2 + \beta (s^6+t^6+u^6) \right) \,.
\end{equation}
After modding out the $u$ required to satisfy the four-point $(n-3)!$ relations we are left with two  independent permutation invariant basis elements of the correct dimension.  Both contribute to abelian-$Z$ theory (and consequently the abelianized open string) at the $\alpha'^8$ order, suggesting that their coefficients may be ultimately fixed by massive mode resonance unitarity considerations of the UV completion~(c.f.~e.g.~ref.~\cite{Caron-Huot:2016icg}).  It would be fascinating if other conditions can be imposed to uniquely fix even these higher $\kappa$ ansatze, but also to understand the structure of these special polynomials.

\section{Uniqueness from UV behavior}
\label{UVBehaviorSection}
\subsection{Doubly-ordered bi-adjoint amplitudes}
\begin{claim}\label{biadjthm}
The  doubly-ordered bi-adjoint amplitudes are fixed uniquely by locality and the following UV single hard scaling $z \,p_{n}\rightarrow \infty$, $\delta(i,[j,k])$:
\begin{numcases}{A_n(\sigma_1|\sigma_2)\sim}
  \mathcal{O}(z^{-3}),& \textrm{for $\delta$ compatible with both $\sigma_1$ and $\sigma_2$}\\
  \mathcal{O}(z^{-2}) ,& \textrm{for $\delta$ compatible with either $\sigma_1$ or $\sigma_2$}
\end{numcases}
\end{claim}
As mentioned before, the amplitudes can have even better scalings in particular situations. However, the above conditions are sufficient to fully determine all bi-adjoint amplitudes. The proof is a much simpler version of the argument applied to Yang-Mills in ref.~\cite{Rodina:2016mbk}. Briefly, we only need to show that the leading soft piece of the bi-adjoint amplitude is fixed uniquely by imposing UV constraints. This is because any term in $\phi^3$ amplitudes, for four point and above, has at least two cubic poles, and therefore shows up in at least one single-soft theorem. Hence the bi-adjoint is completely fixed by just its leading soft theorem, similar to other cases as explained in ref.~\cite{RodinaSoft}. We carry out the proof by induction in Appendix (\ref{adjproof}), for the simplifying case when $\sigma_1=\sigma_2=(1,2,\ldots,n)$, as other configurations follow from identical reasoning. 

Here we examine instead the first step of the induction, that UV constraints fix the five-point amplitude. The check is simple,  but it is instructive, as unlike the examples to follow, the bi-adjoint has trivial numerator structure, and obeys no on-shell constraints other than unitarity (with simple soft theorems as a consequence) and amplitude relations once it is dressed with a single copy of color factors. It therefore most transparently demonstrates that improved UV scaling is directly tied to unitarity. Consider the five-point example for $\sigma_1=\sigma_2=(1,2,3,4,5)$, and take $zp_1$ large, choosing $\delta(3,[4,5])$:
\begin{align}
\label{5p}
A_5=&\frac{a_1}{s_{12}s_{34}}+\frac{a_2}{s_{23}s_{45}}+\frac{a_3}{s_{34}s_{51}}+\frac{a_4}{s_{45}s_{12}}+\frac{a_5}{s_{51}s_{23}}\\
\nonumber \rightarrow &\frac{1}{z^2}\left(\frac{a_1}{p_1.p_2 \left(p_1.p_2+p_1.p_5\right)}+\frac{a_2}{p_1.p_5 \left(p_1.p_2+p_1.p_5\right)}+\frac{a_3}{p_1.p_2 \left(p_1.p_2+p_1.p_4+p_1.p_5\right)}\right.\\
&\left.-\frac{a_4}{p_1.p_2 \left(p_1.p_2+p_1.p_4+p_1.p_5\right)}-\frac{a_5}{p_1.p_2 p_1.p_5}\right)+\mathcal{O}(z^{-3})\\
=&\mathcal{O}(z^{-3})\label{z3} \, .
\end{align}
All 5 terms contribute at order $\mathcal{O}(z^{-2})$ and are needed for the enhancement, meaning that the SHS is probing several cuts at the same time, while any regular unitarity constraint could only probe at most two diagrams at a time. We obtain two constraints: $a_1+a_2-a_5=0$ and $a_3-a_4=0$, with the remaining two  obtained by other hard limits, leading to $a_1=a_2=a_3=a_4=a_5$, fixing the five-point amplitude.

The above example also shows another quite amusing property of the SHS:
it can tell that quantities which apparently look like singly ordered $\phi^3$ partial amplitudes are properly resolved by definition as doubly-ordered bi-adjoint amplitudes.  Consider the simple example of Eqn.~\ref{5p}, which reads like an ordered cubic scalar amplitude. If we make a ``bad'' choice for $\delta(i,[j,k])$, we discover that this object scales as $\mathcal{O}(z^{-1})$, two powers of $z$ worse than (\ref{z3}). Comparing with all the other known singly-ordered partial amplitudes, for which the difference between ``bad'' and ``good'' $\delta(i,[j,k])$ is just one power of $z$, this seems like a puzzle. The discrepancy is easily explained by the rules given by (\ref{biadj}), which seem to suggest we are in fact making a ``doubly'' bad choice. If we are to trust these rules as fundamental, the resolution is obvious: such a scalar amplitude carries an extra hidden identical ordering, so our choice for $\delta(i,[j,k])$ is breaking the rule twice!

Next we move on to more complicated theories, which have extra kinematic structure in the numerators. If the bi-adjoint scalar example seems to imply the UV scalings are perhaps just compactly imposing unitarity, we will see much more is true. As discussed previously, general theories (and in particular EFT's) have contact terms, which are invisible to factorization constraints. Therefore, somehow, UV scalings are imposing other symmetries on the amplitudes, and not just factorization.

\subsection{NLSM}
\begin{claim}
NLSM amplitudes are fixed uniquely by locality, unitarity, $\textrm{[mass dimension]}$ $=2$, and the following UV behaviors:\\
\textbf{A}. Single hard scaling $z \,p_{n}\rightarrow \infty$, with $\delta(i,[j,k])$:
\begin{numcases}{ A_n(\sigma)\sim}\label{nlsmshs1}
  \mathcal{O}(z^0),& \textrm{for $\delta$ compatible with $\sigma$  }\\
 \label{nlsmshs2} \mathcal{O}(z^1),& \textrm{otherwise}
\end{numcases}
\textbf{B}. Two particle shift $[i,j\rangle$ (\ref{bcfw}):
\begin{numcases}
{A_n\sim}
\label{nlsm2s}
  \mathcal{O}(z^1), &\textrm{ for adjacent $i,j$}\\
  \mathcal{O}(z^0), & \textrm{ for non-adjacent $i,j$}
\end{numcases}
\end{claim}
We only need to show that any contact terms cannot independently satisfy the required scaling. Since it is a simple linear combination of kinematic invariants, the proof is straightforward via soft limits and induction. Using $p_{n+1}=\tau \,p_{n+1}$, we expand the contact term at $(n+1)$ around $\tau=0$:
\begin{align}
C_{n+1}=\tau^0 C_{n+1}^{(0)}+\mathcal{O}(\tau) \, .
\end{align}
The leading piece $C_{n+1}^{(0)}$ is simply the lower point ansatz $C_n$, which by assumption cannot satisfy the SHS or 2S constraints. Therefore $C_{n+1}$ scales as $\mathcal{O}(\tau)$ in the single soft limit, and is therefore ruled out by uniqueness from the Adler zero.

We expect that in fact the UV constraint is much more powerful, and that unitarity follows as a consequence. To test this claim, we setup a local ansatz over quartic graphs of the form:
\begin{align}
B_n^{\textrm{NLSM}}=\sum_i \frac{N_i(p^{n-2})}{D_i} \, .
\end{align} 
Possible contact terms are included in the numerators. Imposing either of these enhanced scalings, we have verified analytically that up through 8 points the ansatz is completely fixed. The check can be performed easily at 4 and 6 points, where the ansatze have just 2 and 135 terms respectively, but grows relatively quickly: already at 8 points the relevant ansatz has 18,480 terms.

\subsection{(Special) Galileon}
As a prelude to the special Galileon, we make two observations about the Galileon vertex, and the (general) Galileon amplitude. Because it will be useful later, we mark the following claim:
\begin{conjecture} 
\label{galVconjecture1}
The Galileon vertex is uniquely fixed by [mass dimension] $=[2n-2]$ and the following UV behaviors:\\
\textbf{A}. Single hard scaling (SHS): $V_n\sim \mathcal{O}(z^4)$\\
\textbf{B}. Two particle shift (2S): $V_n\sim \mathcal{O}(z^2)$.\\
\end{conjecture}
We have verified these claims up through $n=7$, using a polynomial ansatz.

Proving the Galileon vertex actually has these scaling is simple. Recalling the definition (\ref{defgal}), we choose $a=n$, and apply a $[1,n\rangle$ shift. Since $p_n$ does not appear in the matrix, the only $z$ contributions can come from $p_1$ in row 1 and column 1. Therefore the maximum power of $z$ in the determinant is 2. For the single limit, extra $z$ contributions can come in the entries of $M$ which are removed via the on-shell constraint, say $p_2.p_3$. There are two such entries, providing therefore two extra powers of $z$, for a total maximum power of 4 for the SHS. 

Going beyond just the contact terms, the full Galileon, given by the general Lagrangian (\ref{gall}), itself curiously follows from imposing both the SHS and 2S scaling and locality, at least up to $n=7$. Unlike the previous cases, here the coefficient of the contact term is not fixed relative to the factorizing piece (since the contact term is a solution), but surprisingly the factorizing piece itself is fixed.

\begin{claim}
Assuming Conjecture (\ref{galVconjecture1}), the special Galileon (sGal) amplitudes are uniquely fixed by locality, unitarity, $\textrm{[mass dimension]}$ $=[2n-2]$, and the single hard limit scaling:
\begin{align}
\label{sgalshs}
A_n^{\textrm{sGal}}\sim \mathcal{O}(z^3) \, .
\end{align}
\end{claim}
As with previous unitarity proofs above, the claim is that the only contact terms which satisfy the scaling do so by cancellation against factorization channels so can be completely fixed in that way.  We proceed to rule out any additional contact terms that would satisfy the scaling in isolation. From the observation made in Section [\ref{shss}], an $\mathcal{O}(z^3)$ SHS scaling of a contact implies at least a $\mathcal{O}(z^3)$ scaling under the hard BCFW double-shift (2S scaling).  This 2S scaling is automatically improved to $\mathcal{O}(z^2)$ by permutation invariance \cite{McGady:2014lqa}, which implies that the contact must uniquely be Galileon by conjecture~\ref{galVconjecture1}.   But via the same conjecture, such a vertex has a $z^4$ SHS, so would violate the specified scaling condition of $z^3$.

We can see what happens if we do not impose unitarity. The ansatz in this case is:
\begin{align}
B_n^{\textrm{sGal}}=\sum_i \frac{N_i(p^{3n-6})}{D_i} \, .
\end{align}
Because of the higher mass dimension of the numerators, only the $n=4$ and $n=6$ cases are straightforward to verify. We find in these cases that the ansatz is fixed, and unitarity emerges from the UV constraints.

The parallel to the IR story is interesting to note. There, the special Galileon was selected by demanding a further improved $\mathcal{O}(\tau^3)$ soft limit, up from the $\mathcal{O}(\tau^2)$ satisfied by the general Galileon. The corresponding scalings in the hard limit are on the other hand $\mathcal{O}(z^4)$ for the Galileon, improved to $\mathcal{O}(z^3)$ for the sGal.

\subsection{Born-Infeld}
\begin{claim}
\label{BIfrom2S}
BI amplitudes are uniquely fixed by locality, unitarity, $\textrm{[mass dimension]}$ $= n$ and a BCFW shift (\ref{vbcfw}) scaling of 
\begin{align}
A_n^{\textrm{BI}}\sim\mathcal{O}(z^0)\, ,
\end{align}
\end{claim}
As before, we only need to show the contact term cannot independently scale, without engagement with terms on factorization channels, as $\mathcal{O}(z^0)$ under BCFW shifts. To do this, we expand in a double soft limit
\begin{align}
C_{n+2}^{\textrm{BI}}=\tau^0 C_{n+2}^{(0)}+\tau C_{n+2}^{(1)}+\ldots \, ,
\end{align}
and show that the BCFW constraint rules out $C^{(0)}$, $C^{(1)}$, and $C^{(2)}$. This is sufficient to rule out the whole term, since it is a polynomial of mass dimension $[n+2]$, and it cannot have a soft limit scaling of $\mathcal{O}(\tau^3)$ in all possible double soft limits, a proof we leave to the appendix (\ref{thmpol}).
In this setup the $[n+1,n+2\rangle$ shift is very constraining. At the first order, demanding $\mathcal{O}(z^0)$ under this shift fixes
\begin{align}
C_{n+2}^{(0)}=\epsilon_{n+1}.\epsilon_{n+2}C_n(p^{n+2}) \, ,
\end{align}
where $C_n(p^{n+2})$ is some $n$-point general polynomial.
Taking advantage of the fact that $p_n$ can be removed from $C_n$ via momentum conservation, we now impose a $[n+2,n\rangle$ shift. Since $C_n$ is linear in $e_n$, the shift produces a piece proportional to $z$, which cannot cancel against anything, and therefore $C_{n+2}^{(0)}$ must vanish. The next orders are slightly more involved but can be fixed using similar arguments, proving that BCFW scaling plus unitarity fixes the ansatz uniquely.

It is already quite surprising that BI can also be fixed simply by its high energy behavior, even without assuming gauge invariance or a Lagrangian form like (\ref{BIform}), but we find evidence that even unitarity may be dropped from the starting assumptions. This is easily verified at 4 points, where the BI amplitude coincides with the numerator of the YM amplitude. However, even at 6 points the ansatz starts to become prohibitively large, containing over $3\times10^6$ terms. To check the conjecture we therefore Taylor expand the ansatz in a double soft limit, and check whether the ansatz is fixed order by order:
\begin{align}
B_{6}\rightarrow \frac{1}{\tau}B_6^{(-1)}+\tau^0 B_6^{(0)}+\tau B_6^{(1)}+\tau^2 B_6^{(2)}+\ldots \, .
\end{align} 
We have indeed verified that imposing the scaling is enough to uniquely fix all terms up to and including $\tau^2$, and according to the arguments of the type given in ref.~\cite{RodinaSoft}, this is sufficient to fully fix the amplitude.

\subsubsection{Supersymmetric DBI-VA}

The UV constraints can be applied to 4D kinematics as well. We find that the photon and fermion sectors of DBI-VA in four dimensions are uniquely fixed by locality, mass dimension, helicity weight, and two particle shift scalings:
\begin{numcases}{A^\textrm{photon}\sim}
\mathcal{O}(z^0) & for $(-,-)$, $(+,+)$ and $(-,+)$\\
\mathcal{O}(z^4)&  for $(+,-)$
\end{numcases}
\begin{numcases}{A^\textrm{fermion}\sim}
\mathcal{O}(z^0) & for $(-,-)$ and $(+,+)$\\
\mathcal{O}(z^1) & for $(-,+)$ \\
\mathcal{O}(z^3)&  for $(+,-)$
\end{numcases}
For DBI-VA, we can write a local 4D ansatz:
\begin{align}
B_n=\sum \frac{N}{D} \, ,
\end{align}
where the N are polynomials of spinor dot products $\langle i,j\rangle$ and $[i,j]$, have mass dimension $[2n-2]$, and a corresponding helicity weight for each particle. We have checked the conjecture up through $n=8$.

The scalar part of this theory, DBI, is a surprising exception in this context. Even though the infrared properties are enough to constrain it, the UV behavior apparently is not. Technically, this is because it is lower mass dimension than the Galileon, but obeys the same UV scalings ($\mathcal{O}(z^2)$ BCFW shift, $\mathcal{O}(z^3)$ SHS).

\subsection{The Double-Copy and UV behavior}
The double copy procedure makes it clear how BI and sGal inherit the SHS UV scalings from YM and NLSM (see also ref.~\cite{Square,Boels:2012sy} for discussions on the usual BCFW shift). We will use the following KLT representation:
\begin{align}
\label{KLT}
M_n=\sum_{\sigma(\alpha),\sigma(\beta)}A(1,\sigma(\alpha),n,n-1)S[\alpha|\beta]A(1,\sigma(\beta),n-1,n)
\end{align} 
where the KLT matrix can be defined recursively as:
\begin{align}
\label{KLTmatrix}
S[A, j | B, j, C]_{i}=\left(k_{i B} \cdot k_{j}\right) S[A | B, C]_{i}, \quad S[\emptyset | \emptyset]_{i} \equiv 1 \, ,
\end{align}
with $k_{iB}\equiv k_{i}+k_{b_{1}}+\cdots+k_{b_{|B|}}$, and we choose $i=1$.

This form is convenient because particles $n$ and $n-1$ are always adjacent on the right-hand side of eq. (\ref{KLT}), and furthermore $p_{n-1}$ and $p_{n}$ do not appear manifestly in the KLT matrix. Therefore, imposing the shift $[n-1,n\rangle$, we find the scalings:
\begin{align}
[\textrm{BI}]&=[\textrm{YM}]+[\textrm{NLSM}]=(-1)+(1)=0\\
[\textrm{sGal}]&=[\textrm{NLSM}]+[\textrm{NLSM}]=2
\end{align}
as expected. 
For the SHS, we can make the scaling manifest with any choice $\delta(p_{n-1},[p_1.p_i])$, taking $n$ hard. Since $n$ and $n-1$ do not appear explicitly, the only $p_n$ contribution can come from eliminating $p_1.p_i$. It is easy to see that any term of the type $p_1.p_i$ appears exactly once in every $S[\alpha|\beta]_1$, contributing one power of $z$.  We therefore obtain a SHS scaling
\begin{align}
[\textrm{sGal}]=[\textrm{NLSM}]+[S_{\textrm{KLT}}]+[\textrm{NLSM}]=3\ ,
\end{align}
as expected. The same argument shows [GR]=[BI]=3.

\section{Summary and outlook}
To summarize, we have considered both color-kinematic and UV constraints on tree-level scattering amplitudes, novelly introducing a single-particle hard shift scaling.  In conjunction with unitarity we proved a number of uniqueness claims, and collected evidence that for many of these theories unitarity could emerge from such constraints alone.  As such results could potentially be far more reaching, suggesting structure that allows unitarity to follow from UV behavior, they are worth summarizing separately.  The following quantities may be uniquely fixed by UV conditions:
\begin{enumerate}
\item  NLSM amplitudes  (verified through 8 points).
\item The Galileon vertex  (verified through 7 points).
\item The special Galileon amplitude  (verified through 6 points).
\item Arbitrary dimension Born-Infeld  (verified through 6 points).
\item 4D Supersymmetric DBI-VA  vector and fermion amplitudes (verified through 8 points).
\end{enumerate}
These results extend the list of cases when amplitudes can be derived solely from new principles. This makes it increasingly plausible that a different formulation exists, where some of these properties are primary, at the expense of manifest factorization and space-time descriptions. This is concordant with the ``Amplituhedron'' program \cite{Arkani-Hamed:2013jha} (and recent generalizations \cite{Arkani-Hamed:2017mur}), where both locality and unitarity follow from more basic geometric principles. 

Since these are all massless theories it is perhaps not surprising that the IR and UV limits actually contain equivalent information. However, a clear way to go from one to the other is still lacking. It would be interesting if the recently discovered conformal symmetry in D-dimensional YM and GR tree amplitudes  plays any role in this context \cite{Loebbert:2018xce}. It is also likely that some symmetry must be behind these high energy limits for EFT's, similar to the ``enhanced spin symmetry'' that was discovered for YM and GR  \cite{ArkaniHamed:2008yf}.  

Since the spectrum of multi-particle theories, unless hard-partitioned into some sort of grassmanian indexed generator (as is frequently done by defining an on-shell  superspace), can often be inaccessible except at the (multi-)loop level it will be interesting to see what kind of constraints one can expect on gauge-invariant components of integrands.  This is a program that has already produced interesting results considering generalized unitarity-cuts~\cite{Bourjaily:2018omh}. Given the intimate relation between supersymmetry and UV behavior, it will not be surprising if UV scaling can reduce the number of cuts required to completely specify supersymmetric gauge theory integrands.  A particularly interesting avenue would be to investigate any barriers to relating such UV scaling in non-supersymmetric theories to their beta functions (c.f~ref.~\cite{Caron-Huot:2016cwu}).

We also note that in the context of celestial amplitudes (cf.~ref.~\cite{Pasterski:2017kqt,Pasterski:2016qvg,Schreiber:2017jsr}), the UV scaling of amplitudes makes an appearance, as does the infrared behavior (cf. ref.~\cite{Donnay:2018neh,Nandan:2019jas,Fan:2019emx,Guevara:2019ypd}).  The Mellin transform is an integral over the energies of the amplitude, and is sensitive to UV divergences, hence string theory completions are needed for consistency even when discussing purely gravitational amplitudes \cite{Stieberger:2018edy}. As the transform mixes the IR and UV of the amplitudes, any possible IR/UV connection may ultimately have an impact on celestial amplitudes as well.

\section{Acknowledgments}
We thank Zvi Bern, Marco Chiodaroli, Henrik Johansson, Karol Kampf, Ian Low, Radu Roiban, Jaroslav Trnka, Ingrid Vazquez-Holm, Congkao Wen,  Zhewei Yin, and Suna Zekioglu 
for many useful and interesting related discussions.
JJMC and LR are supported by the European Research Council under ERC-STG-639729, {\it Strategic Predictions for Quantum
Field Theories\/}. 

\section{Appendix}

\subsection{Bi-adjoint soft theorem from UV scaling}\label{adjproof}
We wish to prove that the leading order soft theorem of bi-adjoint scalar, given by:
\begin{align}
A_{n+1}(1,2,\ldots,n)\rightarrow \left(\frac{1}{p_{n+1}.p_1}+\frac{1}{p_{n+1}.p_n}\right) \, ,
\end{align} 
can be fixed by the SHS scalings of Claim~(\ref{biadjthm}). Since we are only dealing with the case $\sigma_1=\sigma_2$, the constraints simplify, and we can compactly refer to them as:
\begin{align} \label{en} 
E_n\equiv \lim_{z p\rightarrow \infty}A_n\delta(i,[j,k])\sim
\mathcal{O}(z^{-3}), \textrm{$\forall p$, $i$ not adjacent to $p$, and $j>i$, or $k<i$} \, .
\end{align}
As will become clear immediately, we in fact need to prove a somewhat stronger statement for the inductive argument to close: we impose the scalings for all particles, except one we denote $h$. Using our previous notation, further define:
\begin{align}
E_n(h)\equiv E_n, \forall p\neq h \, .
\end{align}
Now we begin the inductive argument, by assuming $A_n(\sigma)$ is fixed uniquely by the constraints $E_n(h)$, $\forall\, h=\overline{1,n}$. We wish to show that $E_{n+1}(h')$ acting on a local ansatz $B_{n+1}$ implies $E_n(h)$ constraints acting on a lower point ansatz $B_n$, which is then fixed. Taking a soft limit:
\begin{align}\label{galsoft}
B_{n+1}(\sigma,n+1)\rightarrow \frac{B_{n;1}(\sigma)}{p_{n+1}.p_1}+ \frac{B_{n;n}(\sigma)}{p_{n+1}.p_n} \, ,
\end{align}
where, because of the factorizing propagator structure, $B_{n;i}$ are two different lower point local ansatze. Now the need to single out a particle $h$ becomes clear. If we were to take $p_1$ (or $p_n$) hard, demanding $\mathcal{O}(z^{-3})$ on $B_{n+1}$, because of the denominators, this would only translate to a constraint $\mathcal{O}(z^{-2})$ on $B_{n;1}$ (or $B_{n;n;}$), and the inductive argument would not close. Even less subtly, once we have taken $n+1$ soft, we cannot of course take it hard, also leading to the induction not closing. We can resolve both these issues by starting from a slightly weaker constraint, where we do not impose the scaling for some particular particle $h$. We merely need to choose $h=n+1$ for the high point ansatz, and $h=1$ $(h=n)$ for $B_{n;1}$ ($B_{n;n}$), solving both problems at once.

It follows quickly that all constraints in $E_{n+1}(n+1)$, applied to (\ref{galsoft}), translate to equivalent constraints $E_n(1)$ (or $E_n(n)$) applied to their respective $B_{n;1}$ (or $B_{n;n})$, except two special choices: taking $p_1$ hard with $\delta(n,a,b)$ or taking $p_n$ hard with $\delta(1,a,b)$, for some $a,b$. For the moment, we just note that since 1 and $n$ are adjacent at $n$-points, and so are in fact not part of $E_n$ as defined by (\ref{en}). Therefore, by assumption we have enough constraints to fix the lower point ansatze, each up to some coefficient:
\begin{align}
B_{n+1}\rightarrow \frac{a_1 A_n}{p_{n+1}.p_1}+ \frac{a_n A_n}{p_{n+1}.p_n} \, .
\end{align}
Finally we fix the remaining freedom with the constraint we just ignored: imposing $p_1$ hard with $\delta(n,a,b)$. This choice mixes the two terms
\begin{align}
\mathcal{O}(z^{-3})\sim B_{n+1}\rightarrow A_n(z p_1)\left(\frac{a_1}{z p_{n+1}.p_1}- \frac{a_n}{zp_{n+1}.p_1}\right) \, .
\end{align}
Note that 1 and $n$ are non-adjacent at $(n+1)$-points, so we demand $B_{n+1}$  scale as $\mathcal{O}(z^{-3})$, whereas by assumption $A_n$ scales as $\mathcal{O}(z^{-1})$ since 1 and $n$ are adjacent at $n$-points. This implies we must have $a_1=a_n$, finally fixing the full leading soft theorem for $B_{n+1}$, and hence $B_{n+1}=A_{n+1}$, completing the induction.

\subsection{BI contact term double soft scaling}\label{thmpol}
We will prove there is no polynomial of mass dimension $[n]$ with double soft scaling $\mathcal{O}(\tau^3)$. Taking a double soft limit in $p_{n+1}=q$ and $p_{n+2}=q$ (which by assumption must start at order $\tau^3$):
\begin{align}
C_{n+2}(p^{n+2})=\tau^3 \left(q^\mu q.p C^{\mu}_n(p^{n-1})+q^\mu q^\nu q^\rho C^{\mu\nu\rho}_n(p^{n-1})+\ldots \right) \, ,
\end{align}
next we impose the other double soft limits. In fact, we will not impose any double soft limit involving particle $n$, which we remove via momentum conservation. This is a stronger statement to prove, but what it buys us is now we do not need to worry about cross term cancellations. Therefore now we need to show there are no {\em tensor} polynomials $C^{\mu}_n(p^{n-1})$ or $C_n^{\mu\nu\rho}(p^{n-1})$ with $\mathcal{O}(\tau^3)$ scaling. We can keep repeating the argument until we end up needing to show that a ``totally tensorized'' polynomial:
\begin{align}
C_n^{\mu_1\ldots\mu_{k}}(p^{k}) \, ,
\end{align}
cannot have an enhanced scaling in arbitrarily high dimensions, which is obvious, as there can be no non-trivial cancellations between terms of this tensor. 

\newpage
\bibliography{UVmethods}

\providecommand{\href}[2]{#2}\begingroup\raggedright\begin{thebibliography}{100}

\bibitem{Arkani-Hamed:2016rak}
N.~Arkani-Hamed, L.~Rodina and J.~Trnka, \emph{{Locality and unitarity of
  scattering amplitudes from singularities and gauge invariance}},
  \href{https://doi.org/10.1103/PhysRevLett.120.231602}{\emph{Phys. Rev. Lett.}
  {\bfseries 120} (2018) 231602}
  [\href{https://arxiv.org/abs/1612.02797}{{\ttfamily 1612.02797}}].

\bibitem{RodinaGaugeInv}
L.~Rodina, \emph{{Uniqueness from gauge invariance and the Adler zero}},
  \href{https://arxiv.org/abs/1612.06342}{{\ttfamily 1612.06342}}.

\bibitem{Susskind:1970gf}
L.~Susskind and G.~Frye, \emph{{Algebraic aspects of pionic duality diagrams}},
  \href{https://doi.org/10.1103/PhysRevD.1.1682}{\emph{Phys. Rev.} {\bfseries
  D1} (1970) 1682}.

\bibitem{Osborn:1969ku}
H.~Osborn, \emph{{Implications of adler zeros for multipion processes}},
  \href{https://doi.org/10.1007/BF02755724}{\emph{Lett. Nuovo Cim.} {\bfseries
  2S1} (1969) 717}.

\bibitem{AdlerZero}
S.~L. Adler, \emph{{Consistency conditions on the strong interactions implied
  by a partially conserved axial vector current}},
  \href{https://doi.org/10.1103/PhysRev.137.B1022}{\emph{Phys. Rev.} {\bfseries
  137} (1965) B1022}.

\bibitem{Ellis:1970nt}
J.~R. Ellis and B.~Renner, \emph{{On the relationship between chiral and dual
  models}}, \href{https://doi.org/10.1016/0550-3213(70)90515-8}{\emph{Nucl.
  Phys.} {\bfseries B21} (1970) 205}.

\bibitem{ArkaniHamed2008gz}
N.~Arkani-Hamed, F.~Cachazo and J.~Kaplan, \emph{{What is the simplest quantum
  field theory?}}, \href{https://doi.org/10.1007/JHEP09(2010)016}{\emph{JHEP}
  {\bfseries 09} (2010) 016} [\href{https://arxiv.org/abs/0808.1446}{{\ttfamily
  0808.1446}}].

\bibitem{Cheung:2014dqa}
C.~Cheung, K.~Kampf, J.~Novotny and J.~Trnka, \emph{{Effective Field Theories
  from Soft Limits of Scattering Amplitudes}},
  \href{https://doi.org/10.1103/PhysRevLett.114.221602}{\emph{Phys. Rev. Lett.}
  {\bfseries 114} (2015) 221602}
  [\href{https://arxiv.org/abs/1412.4095}{{\ttfamily 1412.4095}}].

\bibitem{Cheung:2016drk}
C.~Cheung, K.~Kampf, J.~Novotny, C.-H. Shen and J.~Trnka, \emph{{A periodic
  table of effective field theories}},
  \href{https://doi.org/10.1007/JHEP02(2017)020}{\emph{JHEP} {\bfseries 02}
  (2017) 020} [\href{https://arxiv.org/abs/1611.03137}{{\ttfamily
  1611.03137}}].

\bibitem{Cachazo:2014fwa}
F.~Cachazo and A.~Strominger, \emph{{Evidence for a New Soft Graviton
  Theorem}},  \href{https://arxiv.org/abs/1404.4091}{{\ttfamily 1404.4091}}.

\bibitem{Strominger:2014pwa}
A.~Strominger and A.~Zhiboedov, \emph{{Gravitational Memory, BMS
  Supertranslations and Soft Theorems}},
  \href{https://doi.org/10.1007/JHEP01(2016)086}{\emph{JHEP} {\bfseries 01}
  (2016) 086} [\href{https://arxiv.org/abs/1411.5745}{{\ttfamily 1411.5745}}].

\bibitem{He2014laa}
T.~He, V.~Lysov, P.~Mitra and A.~Strominger, \emph{{BMS supertranslations and
  Weinberg's soft graviton theorem}},
  \href{https://doi.org/10.1007/JHEP05(2015)151}{\emph{JHEP} {\bfseries 05}
  (2015) 151} [\href{https://arxiv.org/abs/1401.7026}{{\ttfamily 1401.7026}}].

\bibitem{Hawking:2016msc}
S.~W. Hawking, M.~J. Perry and A.~Strominger, \emph{{Soft Hair on Black
  Holes}}, \href{https://doi.org/10.1103/PhysRevLett.116.231301}{\emph{Phys.
  Rev. Lett.} {\bfseries 116} (2016) 231301}
  [\href{https://arxiv.org/abs/1601.00921}{{\ttfamily 1601.00921}}].

\bibitem{RodinaSoft}
L.~Rodina, \emph{{Scattering amplitudes from soft theorems and infrared
  behavior}}, \href{https://doi.org/10.1103/PhysRevLett.122.071601}{\emph{Phys.
  Rev. Lett.} {\bfseries 122} (2019) 071601}
  [\href{https://arxiv.org/abs/1807.09738}{{\ttfamily 1807.09738}}].

\bibitem{BCFW}
R.~Britto, F.~Cachazo, B.~Feng and E.~Witten, \emph{{Direct proof of tree-level
  recursion relation in Yang-Mills theory}},
  \href{https://doi.org/10.1103/PhysRevLett.94.181602}{\emph{Phys. Rev. Lett.}
  {\bfseries 94} (2005) 181602}
  [\href{https://arxiv.org/abs/hep-th/0501052}{{\ttfamily hep-th/0501052}}].

\bibitem{Benincasa:2007xk}
P.~Benincasa and F.~Cachazo, \emph{{Consistency Conditions on the S-Matrix of
  Massless Particles}},  \href{https://arxiv.org/abs/0705.4305}{{\ttfamily
  0705.4305}}.

\bibitem{McGady:2013sga}
D.~A. McGady and L.~Rodina, \emph{{Higher-spin massless $S$-matrices in
  four-dimensions}},
  \href{https://doi.org/10.1103/PhysRevD.90.084048}{\emph{Phys. Rev.}
  {\bfseries D90} (2014) 084048}
  [\href{https://arxiv.org/abs/1311.2938}{{\ttfamily 1311.2938}}].

\bibitem{Kampf:2013vha}
K.~Kampf, J.~Novotny and J.~Trnka, \emph{{Tree-level amplitudes in the
  nonlinear sigma model}},
  \href{https://doi.org/10.1007/JHEP05(2013)032}{\emph{JHEP} {\bfseries 05}
  (2013) 032} [\href{https://arxiv.org/abs/1304.3048}{{\ttfamily 1304.3048}}].

\bibitem{Kampf:2012fn}
K.~Kampf, J.~Novotny and J.~Trnka, \emph{{Recursion relations for tree-level
  amplitudes in the $SU(N)$ nonlinear sigma model}},
  \href{https://doi.org/10.1103/PhysRevD.87.081701}{\emph{Phys. Rev.}
  {\bfseries D87} (2013) 081701}
  [\href{https://arxiv.org/abs/1212.5224}{{\ttfamily 1212.5224}}].

\bibitem{Cheung:2015cba}
C.~Cheung, C.-H. Shen and J.~Trnka, \emph{{Simple Recursion Relations for
  General Field Theories}},
  \href{https://doi.org/10.1007/JHEP06(2015)118}{\emph{JHEP} {\bfseries 06}
  (2015) 118} [\href{https://arxiv.org/abs/1502.05057}{{\ttfamily
  1502.05057}}].

\bibitem{Luo:2015tat}
H.~Luo and C.~Wen, \emph{{Recursion relations from soft theorems}},
  \href{https://doi.org/10.1007/JHEP03(2016)088}{\emph{JHEP} {\bfseries 03}
  (2016) 088} [\href{https://arxiv.org/abs/1512.06801}{{\ttfamily
  1512.06801}}].

\bibitem{Cachazo2016njl}
F.~Cachazo, P.~Cha and S.~Mizera, \emph{{Extensions of theories from soft
  limits}}, \href{https://doi.org/10.1007/JHEP06(2016)170}{\emph{JHEP}
  {\bfseries 06} (2016) 170}
  [\href{https://arxiv.org/abs/1604.03893}{{\ttfamily 1604.03893}}].

\bibitem{Low:2019ynd}
I.~Low and Z.~Yin, \emph{{Soft Bootstrap and Effective Field Theories}},
  \href{https://arxiv.org/abs/1904.12859}{{\ttfamily 1904.12859}}.

\bibitem{Elvang:2018dco}
H.~Elvang, M.~Hadjiantonis, C.~R.~T. Jones and S.~Paranjape, \emph{{Soft
  Bootstrap and Supersymmetry}},
  \href{https://doi.org/10.1007/JHEP01(2019)195}{\emph{JHEP} {\bfseries 01}
  (2019) 195} [\href{https://arxiv.org/abs/1806.06079}{{\ttfamily
  1806.06079}}].

\bibitem{ArkaniHamed:2008yf}
N.~Arkani-Hamed and J.~Kaplan, \emph{{On Tree Amplitudes in Gauge Theory and
  Gravity}}, \href{https://doi.org/10.1088/1126-6708/2008/04/076}{\emph{JHEP}
  {\bfseries 04} (2008) 076} [\href{https://arxiv.org/abs/0801.2385}{{\ttfamily
  0801.2385}}].

\bibitem{Rodina:2016mbk}
L.~Rodina, \emph{{Uniqueness from locality and BCFW shifts}},
  \href{https://arxiv.org/abs/1612.03885}{{\ttfamily 1612.03885}}.

\bibitem{BCJ}
Z.~Bern, J.~J.~M. Carrasco and H.~Johansson, \emph{{New relations for
  gauge-theory amplitudes}},
  \href{https://doi.org/10.1103/PhysRevD.78.085011}{\emph{Phys. Rev.}
  {\bfseries D78} (2008) 085011}
  [\href{https://arxiv.org/abs/0805.3993}{{\ttfamily 0805.3993}}].

\bibitem{BCJLoop}
Z.~Bern, J.~J.~M. Carrasco and H.~Johansson, \emph{{Perturbative quantum
  gravity as a double copy of gauge theory}},
  \href{https://doi.org/10.1103/PhysRevLett.105.061602}{\emph{Phys. Rev. Lett.}
  {\bfseries 105} (2010) 061602}
  [\href{https://arxiv.org/abs/1004.0476}{{\ttfamily 1004.0476}}].

\bibitem{KLT}
H.~Kawai, D.~C. Lewellen and S.~H.~H. Tye, \emph{{A relation between tree
  amplitudes of closed and open strings}},
  \href{https://doi.org/10.1016/0550-3213(86)90362-7}{\emph{Nucl. Phys.}
  {\bfseries B269} (1986) 1}.

\bibitem{Cachazo:2013gna}
F.~Cachazo, S.~He and E.~Y. Yuan, \emph{{Scattering equations and
  Kawai-Lewellen-Tye orthogonality}},
  \href{https://doi.org/10.1103/PhysRevD.90.065001}{\emph{Phys. Rev.}
  {\bfseries D90} (2014) 065001}
  [\href{https://arxiv.org/abs/1306.6575}{{\ttfamily 1306.6575}}].

\bibitem{Cachazo:2013hca}
F.~Cachazo, S.~He and E.~Y. Yuan, \emph{{Scattering of massless particles in
  arbitrary dimensions}},
  \href{https://doi.org/10.1103/PhysRevLett.113.171601}{\emph{Phys. Rev. Lett.}
  {\bfseries 113} (2014) 171601}
  [\href{https://arxiv.org/abs/1307.2199}{{\ttfamily 1307.2199}}].

\bibitem{Cachazo:2013iea}
F.~Cachazo, S.~He and E.~Y. Yuan, \emph{{Scattering of massless particles:
  scalars, gluons and gravitons}},
  \href{https://doi.org/10.1007/JHEP07(2014)033}{\emph{JHEP} {\bfseries 07}
  (2014) 033} [\href{https://arxiv.org/abs/1309.0885}{{\ttfamily 1309.0885}}].

\bibitem{Cachazo2014nsa}
F.~Cachazo, S.~He and E.~Y. Yuan, \emph{{Einstein-Yang-Mills scattering
  amplitudes from scattering equations}},
  \href{https://doi.org/10.1007/JHEP01(2015)121}{\emph{JHEP} {\bfseries 01}
  (2015) 121} [\href{https://arxiv.org/abs/1409.8256}{{\ttfamily 1409.8256}}].

\bibitem{Cachazo2014xea}
F.~Cachazo, S.~He and E.~Y. Yuan, \emph{{Scattering equations and matrices:
  From Einstein to Yang-Mills, DBI and NLSM}},
  \href{https://doi.org/10.1007/JHEP07(2015)149}{\emph{JHEP} {\bfseries 07}
  (2015) 149} [\href{https://arxiv.org/abs/1412.3479}{{\ttfamily 1412.3479}}].

\bibitem{He:2016iqi}
S.~He and Y.~Zhang, \emph{{New Formulas for Amplitudes from Higher-Dimensional
  Operators}}, \href{https://doi.org/10.1007/JHEP02(2017)019}{\emph{JHEP}
  {\bfseries 02} (2017) 019}
  [\href{https://arxiv.org/abs/1608.08448}{{\ttfamily 1608.08448}}].

\bibitem{CheungUnifyingRelations}
C.~Cheung, C.-H. Shen and C.~Wen, \emph{{Unifying relations for scattering
  amplitudes}}, \href{https://doi.org/10.1007/JHEP02(2018)095}{\emph{JHEP}
  {\bfseries 02} (2018) 095}
  [\href{https://arxiv.org/abs/1705.03025}{{\ttfamily 1705.03025}}].

\bibitem{Cheung:2017yef}
C.~Cheung, G.~N. Remmen, C.-H. Shen and C.~Wen, \emph{{Pions as gluons in
  higher dimensions}},
  \href{https://doi.org/10.1007/JHEP04(2018)129}{\emph{JHEP} {\bfseries 04}
  (2018) 129} [\href{https://arxiv.org/abs/1709.04932}{{\ttfamily
  1709.04932}}].

\bibitem{GeneralizedDoubleCopy}
Z.~Bern, J.~J. Carrasco, W.-M. Chen, H.~Johansson and R.~Roiban, \emph{{Gravity
  amplitudes as generalized double copies of gauge-theory amplitudes}},
  \href{https://doi.org/10.1103/PhysRevLett.118.181602}{\emph{Phys. Rev. Lett.}
  {\bfseries 118} (2017) 181602}
  [\href{https://arxiv.org/abs/1701.02519}{{\ttfamily 1701.02519}}].

\bibitem{Bern:2018jmv}
Z.~Bern, J.~J. Carrasco, W.-M. Chen, A.~Edison, H.~Johansson, J.~Parra-Martinez
  et~al., \emph{{Ultraviolet Properties of $\mathcal N = 8$ Supergravity at
  Five Loops}}, \href{https://doi.org/10.1103/PhysRevD.98.086021}{\emph{Phys.
  Rev.} {\bfseries D98} (2018) 086021}
  [\href{https://arxiv.org/abs/1804.09311}{{\ttfamily 1804.09311}}].

\bibitem{FourLoopFormFactor}
R.~H. Boels, B.~A. Kniehl, O.~V. Tarasov and G.~Yang, \emph{{Color-kinematic
  duality for form factors}},
  \href{https://doi.org/10.1007/JHEP02(2013)063}{\emph{JHEP} {\bfseries 02}
  (2013) 063} [\href{https://arxiv.org/abs/1211.7028}{{\ttfamily 1211.7028}}].

\bibitem{FiveLoopFormFactor}
G.~Yang, \emph{{Color-kinematics duality and Sudakov form factor at five loops
  for $N=4$ supersymmetric Yang-Mills theory}},
  \href{https://doi.org/10.1103/PhysRevLett.117.271602}{\emph{Phys. Rev. Lett.}
  {\bfseries 117} (2016) 271602}
  [\href{https://arxiv.org/abs/1610.02394}{{\ttfamily 1610.02394}}].

\bibitem{Borsten2013bp}
L.~Borsten, M.~J. Duff, L.~J. Hughes and S.~Nagy, \emph{{Magic square from
  Yang-Mills squared}},
  \href{https://doi.org/10.1103/PhysRevLett.112.131601}{\emph{Phys. Rev. Lett.}
  {\bfseries 112} (2014) 131601}
  [\href{https://arxiv.org/abs/1301.4176}{{\ttfamily 1301.4176}}].

\bibitem{Anastasiou2014qba}
A.~Anastasiou, L.~Borsten, M.~J. Duff, L.~J. Hughes and S.~Nagy,
  \emph{{Yang-Mills origin of gravitational symmetries}},
  \href{https://doi.org/10.1103/PhysRevLett.113.231606}{\emph{Phys. Rev. Lett.}
  {\bfseries 113} (2014) 231606}
  [\href{https://arxiv.org/abs/1408.4434}{{\ttfamily 1408.4434}}].

\bibitem{Nagy:2014jza}
S.~Nagy, \emph{{Chiral squaring}},
  \href{https://doi.org/10.1007/JHEP07(2016)142}{\emph{JHEP} {\bfseries 07}
  (2016) 142} [\href{https://arxiv.org/abs/1412.4750}{{\ttfamily 1412.4750}}].

\bibitem{Anastasiou2015vba}
A.~Anastasiou, L.~Borsten, M.~J. Hughes and S.~Nagy, \emph{{Global symmetries
  of Yang-Mills squared in various dimensions}},
  \href{https://doi.org/10.1007/JHEP01(2016)148}{\emph{JHEP} {\bfseries 01}
  (2016) 148} [\href{https://arxiv.org/abs/1502.05359}{{\ttfamily
  1502.05359}}].

\bibitem{Cardoso2016amd}
G.~Cardoso, S.~Nagy and S.~Nampuri, \emph{{Multi-centered $ \mathcal{N}=2 $ BPS
  black holes: a double copy description}},
  \href{https://doi.org/10.1007/JHEP04(2017)037}{\emph{JHEP} {\bfseries 04}
  (2017) 037} [\href{https://arxiv.org/abs/1611.04409}{{\ttfamily
  1611.04409}}].

\bibitem{Cardoso2016ngt}
G.~L. Cardoso, S.~Nagy and S.~Nampuri, \emph{{A double copy for $ \mathcal{N}=2
  $ supergravity: a linearised tale told on-shell}},
  \href{https://doi.org/10.1007/JHEP10(2016)127}{\emph{JHEP} {\bfseries 10}
  (2016) 127} [\href{https://arxiv.org/abs/1609.05022}{{\ttfamily
  1609.05022}}].

\bibitem{Anastasiou:2018rdx}
A.~Anastasiou, L.~Borsten, M.~J. Duff, S.~Nagy and M.~Zoccali, \emph{{Gravity
  as gauge theory squared: A ghost story}},
  \href{https://doi.org/10.1103/PhysRevLett.121.211601}{\emph{Phys. Rev. Lett.}
  {\bfseries 121} (2018) 211601}
  [\href{https://arxiv.org/abs/1807.02486}{{\ttfamily 1807.02486}}].

\bibitem{NeillRothstein}
D.~Neill and I.~Z. Rothstein, \emph{{Classical space-times from the S matrix}},
  \href{https://doi.org/10.1016/j.nuclphysb.2013.09.007}{\emph{Nucl. Phys.}
  {\bfseries B877} (2013) 177}
  [\href{https://arxiv.org/abs/1304.7263}{{\ttfamily 1304.7263}}].

\bibitem{Monteiro2014cda}
R.~Monteiro, D.~O'Connell and C.~D. White, \emph{{Black holes and the double
  copy}}, \href{https://doi.org/10.1007/JHEP12(2014)056}{\emph{JHEP} {\bfseries
  12} (2014) 056} [\href{https://arxiv.org/abs/1410.0239}{{\ttfamily
  1410.0239}}].

\bibitem{Luna2015paa}
A.~Luna, R.~Monteiro, D.~O'Connell and C.~D. White, \emph{{The classical double
  copy for Taub--NUT spacetime}},
  \href{https://doi.org/10.1016/j.physletb.2015.09.021}{\emph{Phys. Lett.}
  {\bfseries B750} (2015) 272}
  [\href{https://arxiv.org/abs/1507.01869}{{\ttfamily 1507.01869}}].

\bibitem{Ridgway2015fdl}
A.~K. Ridgway and M.~B. Wise, \emph{{Static spherically symmetric Kerr-Schild
  metrics and implications for the classical double copy}},
  \href{https://doi.org/10.1103/PhysRevD.94.044023}{\emph{Phys. Rev.}
  {\bfseries D94} (2016) 044023}
  [\href{https://arxiv.org/abs/1512.02243}{{\ttfamily 1512.02243}}].

\bibitem{Luna2016due}
A.~Luna, R.~Monteiro, I.~Nicholson, D.~O'Connell and C.~D. White, \emph{{The
  double copy: Bremsstrahlung and accelerating black holes}},
  \href{https://doi.org/10.1007/JHEP06(2016)023}{\emph{JHEP} {\bfseries 06}
  (2016) 023} [\href{https://arxiv.org/abs/1603.05737}{{\ttfamily
  1603.05737}}].

\bibitem{Luna2016hge}
A.~Luna, R.~Monteiro, I.~Nicholson, A.~Ochirov, D.~O'Connell, N.~Westerberg
  et~al., \emph{{Perturbative spacetimes from Yang-Mills theory}},
  \href{https://doi.org/10.1007/JHEP04(2017)069}{\emph{JHEP} {\bfseries 04}
  (2017) 069} [\href{https://arxiv.org/abs/1611.07508}{{\ttfamily
  1611.07508}}].

\bibitem{White2016jzc}
C.~D. White, \emph{{Exact solutions for the biadjoint scalar field}},
  \href{https://doi.org/10.1016/j.physletb.2016.10.052}{\emph{Phys. Lett.}
  {\bfseries B763} (2016) 365}
  [\href{https://arxiv.org/abs/1606.04724}{{\ttfamily 1606.04724}}].

\bibitem{Luna:2018dpt}
A.~Luna, R.~Monteiro, I.~Nicholson and D.~O'Connell, \emph{{Type D spacetimes
  and the Weyl double copy}},
  \href{https://arxiv.org/abs/1810.08183}{{\ttfamily 1810.08183}}.

\bibitem{Luna2017dtq}
A.~Luna, I.~Nicholson, D.~O'Connell and C.~D. White, \emph{{Inelastic black
  hole scattering from charged scalar amplitudes}},
  \href{https://doi.org/10.1007/JHEP03(2018)044}{\emph{JHEP} {\bfseries 03}
  (2018) 044} [\href{https://arxiv.org/abs/1711.03901}{{\ttfamily
  1711.03901}}].

\bibitem{Goldberger2017frp}
W.~D. Goldberger, S.~G. Prabhu and J.~O. Thompson, \emph{{Classical gluon and
  graviton radiation from the bi-adjoint scalar double copy}},
  \href{https://doi.org/10.1103/PhysRevD.96.065009}{\emph{Phys. Rev.}
  {\bfseries D96} (2017) 065009}
  [\href{https://arxiv.org/abs/1705.09263}{{\ttfamily 1705.09263}}].

\bibitem{Goldberger2016iau}
W.~D. Goldberger and A.~K. Ridgway, \emph{{Radiation and the classical double
  copy for color charges}},
  \href{https://doi.org/10.1103/PhysRevD.95.125010}{\emph{Phys. Rev.}
  {\bfseries D95} (2017) 125010}
  [\href{https://arxiv.org/abs/1611.03493}{{\ttfamily 1611.03493}}].

\bibitem{Goldberger2017vcg}
W.~D. Goldberger and A.~K. Ridgway, \emph{{Bound states and the classical
  double copy}}, \href{https://doi.org/10.1103/PhysRevD.97.085019}{\emph{Phys.
  Rev.} {\bfseries D97} (2018) 085019}
  [\href{https://arxiv.org/abs/1711.09493}{{\ttfamily 1711.09493}}].

\bibitem{CarrilloGonzalez2017iyj}
M.~Carrillo-Gonz{\'a}lez, R.~Penco and M.~Trodden, \emph{{The classical double
  copy in maximally symmetric spacetimes}},
  \href{https://doi.org/10.1007/JHEP04(2018)028}{\emph{JHEP} {\bfseries 04}
  (2018) 028} [\href{https://arxiv.org/abs/1711.01296}{{\ttfamily
  1711.01296}}].

\bibitem{Gurses:2018ckx}
M.~Gurses and B.~Tekin, \emph{{Classical double copy: Kerr-Schild-Kundt metrics
  from Yang-Mills theory}},
  \href{https://doi.org/10.1103/PhysRevD.98.126017}{\emph{Phys. Rev.}
  {\bfseries D98} (2018) 126017}
  [\href{https://arxiv.org/abs/1810.03411}{{\ttfamily 1810.03411}}].

\bibitem{3PM}
Z.~Bern, C.~Cheung, R.~Roiban, C.-H. Shen, M.~P. Solon and M.~Zeng,
  \emph{{Scattering amplitudes and the conservative Hamiltonian for binary
  systems at third post-Minkowskian order}},
  \href{https://arxiv.org/abs/1901.04424}{{\ttfamily 1901.04424}}.

\bibitem{Bern:2019crd}
Z.~Bern, C.~Cheung, R.~Roiban, C.-H. Shen, M.~P. Solon and M.~Zeng,
  \emph{{Black Hole Binary Dynamics from the Double Copy and Effective
  Theory}},  \href{https://arxiv.org/abs/1908.01493}{{\ttfamily 1908.01493}}.

\bibitem{bcjReview}
Z.~Bern, J.~J. Carrasco, M.~Chiodaroli, H.~Johansson and R.~Roiban, \emph{{The
  Duality Between Color and Kinematics and its Applications}},
  \href{https://arxiv.org/abs/1909.01358}{{\ttfamily 1909.01358}}.

\bibitem{Bern1999bx}
Z.~Bern, A.~De~Freitas and H.~L. Wong, \emph{{On the coupling of gravitons to
  matter}}, \href{https://doi.org/10.1103/PhysRevLett.84.3531}{\emph{Phys. Rev.
  Lett.} {\bfseries 84} (2000) 3531}
  [\href{https://arxiv.org/abs/hep-th/9912033}{{\ttfamily hep-th/9912033}}].

\bibitem{Du2011js}
Y.-J. Du, B.~Feng and C.-H. Fu, \emph{{BCJ relation of color scalar theory and
  KLT relation of gauge theory}},
  \href{https://doi.org/10.1007/JHEP08(2011)129}{\emph{JHEP} {\bfseries 08}
  (2011) 129} [\href{https://arxiv.org/abs/1105.3503}{{\ttfamily 1105.3503}}].

\bibitem{OConnellAlgebras}
N.~E.~J. Bjerrum-Bohr, P.~H. Damgaard, R.~Monteiro and D.~O'Connell,
  \emph{{Algebras for amplitudes}},
  \href{https://doi.org/10.1007/JHEP06(2012)061}{\emph{JHEP} {\bfseries 06}
  (2012) 061} [\href{https://arxiv.org/abs/1203.0944}{{\ttfamily 1203.0944}}].

\bibitem{Chiodaroli2014xia}
M.~Chiodaroli, M.~G{\"u}naydin, H.~Johansson and R.~Roiban, \emph{{Scattering
  amplitudes in $ \mathcal{N}=2 $ Maxwell-Einstein and Yang-Mills/Einstein
  supergravity}}, \href{https://doi.org/10.1007/JHEP01(2015)081}{\emph{JHEP}
  {\bfseries 01} (2015) 081} [\href{https://arxiv.org/abs/1408.0764}{{\ttfamily
  1408.0764}}].

\bibitem{Chiodaroli2015rdg}
M.~Chiodaroli, M.~Gunaydin, H.~Johansson and R.~Roiban, \emph{{Spontaneously
  broken Yang-Mills-Einstein supergravities as double copies}},
  \href{https://doi.org/10.1007/JHEP06(2017)064}{\emph{JHEP} {\bfseries 06}
  (2017) 064} [\href{https://arxiv.org/abs/1511.01740}{{\ttfamily
  1511.01740}}].

\bibitem{Low:2017mlh}
I.~Low and Z.~Yin, \emph{{Ward Identity and Scattering Amplitudes for Nonlinear
  Sigma Models}},
  \href{https://doi.org/10.1103/PhysRevLett.120.061601}{\emph{Phys. Rev. Lett.}
  {\bfseries 120} (2018) 061601}
  [\href{https://arxiv.org/abs/1709.08639}{{\ttfamily 1709.08639}}].

\bibitem{DixonMaltoni}
V.~Del~Duca, L.~J. Dixon and F.~Maltoni, \emph{{New color decompositions for
  gauge amplitudes at tree and loop level}},
  \href{https://doi.org/10.1016/S0550-3213(99)00809-3}{\emph{Nucl. Phys.}
  {\bfseries B571} (2000) 51}
  [\href{https://arxiv.org/abs/hep-ph/9910563}{{\ttfamily hep-ph/9910563}}].

\bibitem{Mizera:2016jhj}
S.~Mizera, \emph{{Inverse of the string theory KLT kernel}},
  \href{https://doi.org/10.1007/JHEP06(2017)084}{\emph{JHEP} {\bfseries 06}
  (2017) 084} [\href{https://arxiv.org/abs/1610.04230}{{\ttfamily
  1610.04230}}].

\bibitem{BjerrumBohr2010ta}
N.~E.~J. Bjerrum-Bohr, P.~H. Damgaard, B.~Feng and T.~Sondergaard,
  \emph{{Gravity and Yang-Mills amplitude relations}},
  \href{https://doi.org/10.1103/PhysRevD.82.107702}{\emph{Phys. Rev.}
  {\bfseries D82} (2010) 107702}
  [\href{https://arxiv.org/abs/1005.4367}{{\ttfamily 1005.4367}}].

\bibitem{Cronin:1967jq}
J.~A. Cronin, \emph{{Phenomenological model of strong and weak interactions in
  chiral U(3) x U(3)}},
  \href{https://doi.org/10.1103/PhysRev.161.1483}{\emph{Phys. Rev.} {\bfseries
  161} (1967) 1483}.

\bibitem{Weinberg:1966fm}
S.~Weinberg, \emph{{Dynamical approach to current algebra}},
  \href{https://doi.org/10.1103/PhysRevLett.18.188}{\emph{Phys. Rev. Lett.}
  {\bfseries 18} (1967) 188}.

\bibitem{Weinberg:1968de}
S.~Weinberg, \emph{{Nonlinear realizations of chiral symmetry}},
  \href{https://doi.org/10.1103/PhysRev.166.1568}{\emph{Phys. Rev.} {\bfseries
  166} (1968) 1568}.

\bibitem{macfarlane1968}
A.~J. MacFarlane, A.~Sudbery and P.~H. Weisz, \emph{On gell-mann's
  $\lambda$-matrices, $d$- and $f$-tensors, octets, and parametrizations of
  $su(3)$}, {\emph{Comm. Math. Phys.} {\bfseries 11} (1968) 77}.

\bibitem{Carrasco2016ldy}
J.~J.~M. Carrasco, C.~R. Mafra and O.~Schlotterer, \emph{{Abelian Z-theory:
  NLSM amplitudes and $\alpha$'-corrections from the open string}},
  \href{https://doi.org/10.1007/JHEP06(2017)093}{\emph{JHEP} {\bfseries 06}
  (2017) 093} [\href{https://arxiv.org/abs/1608.02569}{{\ttfamily
  1608.02569}}].

\bibitem{LowSoft}
I.~Low and Z.~Yin, \emph{{The infrared structure of Nambu-Goldstone bosons}},
  \href{https://doi.org/10.1007/JHEP10(2018)078}{\emph{JHEP} {\bfseries 10}
  (2018) 078} [\href{https://arxiv.org/abs/1804.08629}{{\ttfamily
  1804.08629}}].

\bibitem{Bagger:1996wp}
J.~Bagger and A.~Galperin, \emph{{A New Goldstone multiplet for partially
  broken supersymmetry}},
  \href{https://doi.org/10.1103/PhysRevD.55.1091}{\emph{Phys. Rev.} {\bfseries
  D55} (1997) 1091} [\href{https://arxiv.org/abs/hep-th/9608177}{{\ttfamily
  hep-th/9608177}}].

\bibitem{Bergshoeff:2013pia}
E.~Bergshoeff, F.~Coomans, R.~Kallosh, C.~S. Shahbazi and A.~Van~Proeyen,
  \emph{{Dirac-Born-Infeld-Volkov-Akulov and deformation of supersymmetry}},
  \href{https://doi.org/10.1007/JHEP08(2013)100}{\emph{JHEP} {\bfseries 08}
  (2013) 100} [\href{https://arxiv.org/abs/1303.5662}{{\ttfamily 1303.5662}}].

\bibitem{CheungSoft}
C.~Cheung, K.~Kampf, J.~Novotny, C.-H. Shen, J.~Trnka and C.~Wen, \emph{{Vector
  effective field theories from soft limits}},
  \href{https://doi.org/10.1103/PhysRevLett.120.261602}{\emph{Phys. Rev. Lett.}
  {\bfseries 120} (2018) 261602}
  [\href{https://arxiv.org/abs/1801.01496}{{\ttfamily 1801.01496}}].

\bibitem{Dvali:2000hr}
G.~R. Dvali, G.~Gabadadze and M.~Porrati, \emph{{4-D gravity on a brane in 5-D
  Minkowski space}},
  \href{https://doi.org/10.1016/S0370-2693(00)00669-9}{\emph{Phys. Lett.}
  {\bfseries B485} (2000) 208}
  [\href{https://arxiv.org/abs/hep-th/0005016}{{\ttfamily hep-th/0005016}}].

\bibitem{Nicolis:2008in}
A.~Nicolis, R.~Rattazzi and E.~Trincherini, \emph{{The Galileon as a local
  modification of gravity}},
  \href{https://doi.org/10.1103/PhysRevD.79.064036}{\emph{Phys. Rev.}
  {\bfseries D79} (2009) 064036}
  [\href{https://arxiv.org/abs/0811.2197}{{\ttfamily 0811.2197}}].

\bibitem{deRham:2010kj}
C.~de~Rham, G.~Gabadadze and A.~J. Tolley, \emph{{Resummation of Massive
  Gravity}}, \href{https://doi.org/10.1103/PhysRevLett.106.231101}{\emph{Phys.
  Rev. Lett.} {\bfseries 106} (2011) 231101}
  [\href{https://arxiv.org/abs/1011.1232}{{\ttfamily 1011.1232}}].

\bibitem{Kampf:2014rka}
K.~Kampf and J.~Novotny, \emph{{Unification of Galileon Dualities}},
  \href{https://doi.org/10.1007/JHEP10(2014)006}{\emph{JHEP} {\bfseries 10}
  (2014) 006} [\href{https://arxiv.org/abs/1403.6813}{{\ttfamily 1403.6813}}].

\bibitem{Hinterbichler:2015pqa}
K.~Hinterbichler and A.~Joyce, \emph{{Hidden symmetry of the Galileon}},
  \href{https://doi.org/10.1103/PhysRevD.92.023503}{\emph{Phys. Rev.}
  {\bfseries D92} (2015) 023503}
  [\href{https://arxiv.org/abs/1501.07600}{{\ttfamily 1501.07600}}].

\bibitem{Novotny:2016jkh}
J.~Novotny, \emph{{Geometry of special Galileons}},
  \href{https://doi.org/10.1103/PhysRevD.95.065019}{\emph{Phys. Rev.}
  {\bfseries D95} (2017) 065019}
  [\href{https://arxiv.org/abs/1612.01738}{{\ttfamily 1612.01738}}].

\bibitem{He:2014bga}
S.~He, Y.-t. Huang and C.~Wen, \emph{{Loop Corrections to Soft Theorems in
  Gauge Theories and Gravity}},
  \href{https://doi.org/10.1007/JHEP12(2014)115}{\emph{JHEP} {\bfseries 12}
  (2014) 115} [\href{https://arxiv.org/abs/1405.1410}{{\ttfamily 1405.1410}}].

\bibitem{He:2016vfi}
S.~He, Z.~Liu and J.-B. Wu, \emph{{Scattering equations, twistor-string
  formulas and double-soft limits in four dimensions}},
  \href{https://doi.org/10.1007/JHEP07(2016)060}{\emph{JHEP} {\bfseries 07}
  (2016) 060} [\href{https://arxiv.org/abs/1604.02834}{{\ttfamily
  1604.02834}}].

\bibitem{Guerrieri:2017ujb}
A.~L. Guerrieri, Y.-t. Huang, Z.~Li and C.~Wen, \emph{{On the exactness of soft
  theorems}}, \href{https://doi.org/10.1007/JHEP12(2017)052}{\emph{JHEP}
  {\bfseries 12} (2017) 052}
  [\href{https://arxiv.org/abs/1705.10078}{{\ttfamily 1705.10078}}].

\bibitem{Bern:2014vva}
Z.~Bern, S.~Davies, P.~Di~Vecchia and J.~Nohle, \emph{{Low-energy behavior of
  gluons and gravitons from gauge invariance}},
  \href{https://doi.org/10.1103/PhysRevD.90.084035}{\emph{Phys. Rev.}
  {\bfseries D90} (2014) 084035}
  [\href{https://arxiv.org/abs/1406.6987}{{\ttfamily 1406.6987}}].

\bibitem{Bern:2014oka}
Z.~Bern, S.~Davies and J.~Nohle, \emph{{On Loop Corrections to Subleading Soft
  Behavior of Gluons and Gravitons}},
  \href{https://doi.org/10.1103/PhysRevD.90.085015}{\emph{Phys. Rev.}
  {\bfseries D90} (2014) 085015}
  [\href{https://arxiv.org/abs/1405.1015}{{\ttfamily 1405.1015}}].

\bibitem{Huang2015sla}
Y.-t. Huang and C.~Wen, \emph{{Soft theorems from anomalous symmetries}},
  \href{https://doi.org/10.1007/JHEP12(2015)143}{\emph{JHEP} {\bfseries 12}
  (2015) 143} [\href{https://arxiv.org/abs/1509.07840}{{\ttfamily
  1509.07840}}].

\bibitem{Chen:2014xoa}
W.-M. Chen, Y.-t. Huang and C.~Wen, \emph{{New Fermionic Soft Theorems for
  Supergravity Amplitudes}},
  \href{https://doi.org/10.1103/PhysRevLett.115.021603}{\emph{Phys. Rev. Lett.}
  {\bfseries 115} (2015) 021603}
  [\href{https://arxiv.org/abs/1412.1809}{{\ttfamily 1412.1809}}].

\bibitem{Low:2015ogb}
I.~Low, \emph{{Double Soft Theorems and Shift Symmetry in Nonlinear Sigma
  Models}}, \href{https://doi.org/10.1103/PhysRevD.93.045032}{\emph{Phys. Rev.}
  {\bfseries D93} (2016) 045032}
  [\href{https://arxiv.org/abs/1512.01232}{{\ttfamily 1512.01232}}].

\bibitem{Elvang:2016qvq}
H.~Elvang, C.~R.~T. Jones and S.~G. Naculich, \emph{{Soft photon and graviton
  theorems in effective field theory}},
  \href{https://doi.org/10.1103/PhysRevLett.118.231601}{\emph{Phys. Rev. Lett.}
  {\bfseries 118} (2017) 231601}
  [\href{https://arxiv.org/abs/1611.07534}{{\ttfamily 1611.07534}}].

\bibitem{DiVecchia:2015jaq}
P.~Di~Vecchia, R.~Marotta, M.~Mojaza and J.~Nohle, \emph{{New soft theorems for
  the gravity dilaton and the Nambu-Goldstone dilaton at subsubleading order}},
  \href{https://doi.org/10.1103/PhysRevD.93.085015}{\emph{Phys. Rev.}
  {\bfseries D93} (2016) 085015}
  [\href{https://arxiv.org/abs/1512.03316}{{\ttfamily 1512.03316}}].

\bibitem{DiVecchia:2015oba}
P.~Di~Vecchia, R.~Marotta and M.~Mojaza, \emph{{Soft theorem for the graviton,
  dilaton and the Kalb-Ramond field in the bosonic string}},
  \href{https://doi.org/10.1007/JHEP05(2015)137}{\emph{JHEP} {\bfseries 05}
  (2015) 137} [\href{https://arxiv.org/abs/1502.05258}{{\ttfamily
  1502.05258}}].

\bibitem{Schwab:2014xua}
B.~U.~W. Schwab and A.~Volovich, \emph{{Subleading Soft Theorem in Arbitrary
  Dimensions from Scattering Equations}},
  \href{https://doi.org/10.1103/PhysRevLett.113.101601}{\emph{Phys. Rev. Lett.}
  {\bfseries 113} (2014) 101601}
  [\href{https://arxiv.org/abs/1404.7749}{{\ttfamily 1404.7749}}].

\bibitem{Strominger:2017zoo}
A.~Strominger, \emph{{Lectures on the Infrared Structure of Gravity and Gauge
  Theory}},  \href{https://arxiv.org/abs/1703.05448}{{\ttfamily 1703.05448}}.

\bibitem{Cachazo:2015ksa}
F.~Cachazo, S.~He and E.~Y. Yuan, \emph{{New Double Soft Emission Theorems}},
  \href{https://doi.org/10.1103/PhysRevD.92.065030}{\emph{Phys. Rev.}
  {\bfseries D92} (2015) 065030}
  [\href{https://arxiv.org/abs/1503.04816}{{\ttfamily 1503.04816}}].

\bibitem{He:2018svj}
S.~He and Q.~Yang, \emph{{An Etude on Recursion Relations and Triangulations}},
  \href{https://doi.org/10.1007/JHEP05(2019)040}{\emph{JHEP} {\bfseries 05}
  (2019) 040} [\href{https://arxiv.org/abs/1810.08508}{{\ttfamily
  1810.08508}}].

\bibitem{Benincasa:2007qj}
P.~Benincasa, C.~Boucher-Veronneau and F.~Cachazo, \emph{{Taming Tree
  Amplitudes In General Relativity}},
  \href{https://doi.org/10.1088/1126-6708/2007/11/057}{\emph{JHEP} {\bfseries
  11} (2007) 057} [\href{https://arxiv.org/abs/hep-th/0702032}{{\ttfamily
  hep-th/0702032}}].

\bibitem{Schuster:2008nh}
P.~C. Schuster and N.~Toro, \emph{{Constructing the Tree-Level Yang-Mills
  S-Matrix Using Complex Factorization}},
  \href{https://doi.org/10.1088/1126-6708/2009/06/079}{\emph{JHEP} {\bfseries
  06} (2009) 079} [\href{https://arxiv.org/abs/0811.3207}{{\ttfamily
  0811.3207}}].

\bibitem{Boels:2016xhc}
R.~H. Boels and R.~Medina, \emph{{Graviton and gluon scattering from first
  principles}},
  \href{https://doi.org/10.1103/PhysRevLett.118.061602}{\emph{Phys. Rev. Lett.}
  {\bfseries 118} (2017) 061602}
  [\href{https://arxiv.org/abs/1607.08246}{{\ttfamily 1607.08246}}].

\bibitem{CSW}
F.~Cachazo, P.~Svrcek and E.~Witten, \emph{{MHV vertices and tree amplitudes in
  gauge theory}},
  \href{https://doi.org/10.1088/1126-6708/2004/09/006}{\emph{JHEP} {\bfseries
  09} (2004) 006} [\href{https://arxiv.org/abs/hep-th/0403047}{{\ttfamily
  hep-th/0403047}}].

\bibitem{Berends:1987me}
F.~A. Berends and W.~T. Giele, \emph{{Recursive calculations for processes with
  $n$ gluons}}, \href{https://doi.org/10.1016/0550-3213(88)90442-7}{\emph{Nucl.
  Phys.} {\bfseries B306} (1988) 759}.

\bibitem{Lee:2015upy}
S.~Lee, C.~R. Mafra and O.~Schlotterer, \emph{{Non-linear gauge transformations
  in $D=10$ SYM theory and the BCJ duality}},
  \href{https://doi.org/10.1007/JHEP03(2016)090}{\emph{JHEP} {\bfseries 03}
  (2016) 090} [\href{https://arxiv.org/abs/1510.08843}{{\ttfamily
  1510.08843}}].

\bibitem{Mafra:2016ltu}
C.~R. Mafra, \emph{{Berends-Giele recursion for double-color-ordered
  amplitudes}}, \href{https://doi.org/10.1007/JHEP07(2016)080}{\emph{JHEP}
  {\bfseries 07} (2016) 080}
  [\href{https://arxiv.org/abs/1603.09731}{{\ttfamily 1603.09731}}].

\bibitem{Mizera:2018jbh}
S.~Mizera and B.~Skrzypek, \emph{{Perturbiner methods for effective field
  theories and the double copy}},
  \href{https://doi.org/10.1007/JHEP10(2018)018}{\emph{JHEP} {\bfseries 10}
  (2018) 018} [\href{https://arxiv.org/abs/1809.02096}{{\ttfamily
  1809.02096}}].

\bibitem{Broedel:2014fsa}
J.~Broedel, M.~de~Leeuw, J.~Plefka and M.~Rosso, \emph{{Constraining subleading
  soft gluon and graviton theorems}},
  \href{https://doi.org/10.1103/PhysRevD.90.065024}{\emph{Phys. Rev.}
  {\bfseries D90} (2014) 065024}
  [\href{https://arxiv.org/abs/1406.6574}{{\ttfamily 1406.6574}}].

\bibitem{Froissart:1961ux}
M.~Froissart, \emph{{Asymptotic behavior and subtractions in the Mandelstam
  representation}}, \href{https://doi.org/10.1103/PhysRev.123.1053}{\emph{Phys.
  Rev.} {\bfseries 123} (1961) 1053}.

\bibitem{KleissKuijf}
R.~Kleiss and H.~Kuijf, \emph{{Multi-gluon cross-sections and five jet
  production at hadron colliders}},
  \href{https://doi.org/10.1016/0550-3213(89)90574-9}{\emph{Nucl. Phys.}
  {\bfseries B312} (1989) 616}.

\bibitem{Chen2013fya}
G.~Chen and Y.-J. Du, \emph{{Amplitude relations in non-linear sigma model}},
  \href{https://doi.org/10.1007/JHEP01(2014)061}{\emph{JHEP} {\bfseries 01}
  (2014) 061} [\href{https://arxiv.org/abs/1311.1133}{{\ttfamily 1311.1133}}].

\bibitem{amplituderelationProof}
B.~Feng, R.~Huang and Y.~Jia, \emph{{Gauge amplitude identities by on-shell
  recursion relation in S-matrix program}},
  \href{https://doi.org/10.1016/j.physletb.2010.11.011}{\emph{Phys. Lett.}
  {\bfseries B695} (2011) 350}
  [\href{https://arxiv.org/abs/1004.3417}{{\ttfamily 1004.3417}}].

\bibitem{Carrasco2016ygv}
J.~J.~M. Carrasco, C.~R. Mafra and O.~Schlotterer, \emph{{Semi-abelian
  Z-theory: NLSM$+\phi^{3}$ from the open string}},
  \href{https://doi.org/10.1007/JHEP08(2017)135}{\emph{JHEP} {\bfseries 08}
  (2017) 135} [\href{https://arxiv.org/abs/1612.06446}{{\ttfamily
  1612.06446}}].

\bibitem{Caron-Huot:2016icg}
S.~Caron-Huot, Z.~Komargodski, A.~Sever and A.~Zhiboedov, \emph{{Strings from
  Massive Higher Spins: The Asymptotic Uniqueness of the Veneziano Amplitude}},
  \href{https://doi.org/10.1007/JHEP10(2017)026}{\emph{JHEP} {\bfseries 10}
  (2017) 026} [\href{https://arxiv.org/abs/1607.04253}{{\ttfamily
  1607.04253}}].

\bibitem{McGady:2014lqa}
D.~A. McGady and L.~Rodina, \emph{{Recursion relations for graviton scattering
  amplitudes from Bose symmetry and bonus scaling laws}},
  \href{https://doi.org/10.1103/PhysRevD.91.105010}{\emph{Phys. Rev.}
  {\bfseries D91} (2015) 105010}
  [\href{https://arxiv.org/abs/1408.5125}{{\ttfamily 1408.5125}}].

\bibitem{Square}
Z.~Bern, T.~Dennen, Y.-t. Huang and M.~Kiermaier, \emph{{Gravity as the square
  of gauge theory}},
  \href{https://doi.org/10.1103/PhysRevD.82.065003}{\emph{Phys. Rev.}
  {\bfseries D82} (2010) 065003}
  [\href{https://arxiv.org/abs/1004.0693}{{\ttfamily 1004.0693}}].

\bibitem{Boels:2012sy}
R.~H. Boels and R.~S. Isermann, \emph{{On powercounting in perturbative quantum
  gravity theories through color-kinematic duality}},
  \href{https://doi.org/10.1007/JHEP06(2013)017}{\emph{JHEP} {\bfseries 06}
  (2013) 017} [\href{https://arxiv.org/abs/1212.3473}{{\ttfamily 1212.3473}}].

\bibitem{Arkani-Hamed:2013jha}
N.~Arkani-Hamed and J.~Trnka, \emph{{The amplituhedron}},
  \href{https://doi.org/10.1007/JHEP10(2014)030}{\emph{JHEP} {\bfseries 10}
  (2014) 030} [\href{https://arxiv.org/abs/1312.2007}{{\ttfamily 1312.2007}}].

\bibitem{Arkani-Hamed:2017mur}
N.~Arkani-Hamed, Y.~Bai, S.~He and G.~Yan, \emph{{Scattering forms and the
  positive geometry of kinematics, color and the worldsheet}},
  \href{https://doi.org/10.1007/JHEP05(2018)096}{\emph{JHEP} {\bfseries 05}
  (2018) 096} [\href{https://arxiv.org/abs/1711.09102}{{\ttfamily
  1711.09102}}].

\bibitem{Loebbert:2018xce}
F.~Loebbert, M.~Mojaza and J.~Plefka, \emph{{Hidden conformal symmetry in
  tree-level graviton scattering}},
  \href{https://doi.org/10.1007/JHEP05(2018)208}{\emph{JHEP} {\bfseries 05}
  (2018) 208} [\href{https://arxiv.org/abs/1802.05999}{{\ttfamily
  1802.05999}}].

\bibitem{Bourjaily:2018omh}
J.~L. Bourjaily, E.~Herrmann and J.~Trnka, \emph{{Maximally supersymmetric
  amplitudes at infinite loop momentum}},
  \href{https://doi.org/10.1103/PhysRevD.99.066006}{\emph{Phys. Rev.}
  {\bfseries D99} (2019) 066006}
  [\href{https://arxiv.org/abs/1812.11185}{{\ttfamily 1812.11185}}].

\bibitem{Caron-Huot:2016cwu}
S.~Caron-Huot and M.~Wilhelm, \emph{{Renormalization group coefficients and the
  S-matrix}}, \href{https://doi.org/10.1007/JHEP12(2016)010}{\emph{JHEP}
  {\bfseries 12} (2016) 010}
  [\href{https://arxiv.org/abs/1607.06448}{{\ttfamily 1607.06448}}].

\bibitem{Pasterski:2017kqt}
S.~Pasterski and S.-H. Shao, \emph{{Conformal basis for flat space
  amplitudes}}, \href{https://doi.org/10.1103/PhysRevD.96.065022}{\emph{Phys.
  Rev.} {\bfseries D96} (2017) 065022}
  [\href{https://arxiv.org/abs/1705.01027}{{\ttfamily 1705.01027}}].

\bibitem{Pasterski:2016qvg}
S.~Pasterski, S.-H. Shao and A.~Strominger, \emph{{Flat Space Amplitudes and
  Conformal Symmetry of the Celestial Sphere}},
  \href{https://doi.org/10.1103/PhysRevD.96.065026}{\emph{Phys. Rev.}
  {\bfseries D96} (2017) 065026}
  [\href{https://arxiv.org/abs/1701.00049}{{\ttfamily 1701.00049}}].

\bibitem{Schreiber:2017jsr}
A.~Schreiber, A.~Volovich and M.~Zlotnikov, \emph{{Tree-level gluon amplitudes
  on the celestial sphere}},
  \href{https://doi.org/10.1016/j.physletb.2018.04.010}{\emph{Phys. Lett.}
  {\bfseries B781} (2018) 349}
  [\href{https://arxiv.org/abs/1711.08435}{{\ttfamily 1711.08435}}].

\bibitem{Donnay:2018neh}
L.~Donnay, A.~Puhm and A.~Strominger, \emph{{Conformally Soft Photons and
  Gravitons}}, \href{https://doi.org/10.1007/JHEP01(2019)184}{\emph{JHEP}
  {\bfseries 01} (2019) 184}
  [\href{https://arxiv.org/abs/1810.05219}{{\ttfamily 1810.05219}}].

\bibitem{Nandan:2019jas}
D.~Nandan, A.~Schreiber, A.~Volovich and M.~Zlotnikov, \emph{{Celestial
  Amplitudes: Conformal Partial Waves and Soft Limits}},
  \href{https://arxiv.org/abs/1904.10940}{{\ttfamily 1904.10940}}.

\bibitem{Fan:2019emx}
W.~Fan, A.~Fotopoulos and T.~R. Taylor, \emph{{Soft Limits of Yang-Mills
  Amplitudes and Conformal Correlators}},
  \href{https://doi.org/10.1007/JHEP05(2019)121}{\emph{JHEP} {\bfseries 05}
  (2019) 121} [\href{https://arxiv.org/abs/1903.01676}{{\ttfamily
  1903.01676}}].

\bibitem{Guevara:2019ypd}
A.~Guevara, \emph{{Notes on Conformal Soft Theorems and Recursion Relations in
  Gravity}},  \href{https://arxiv.org/abs/1906.07810}{{\ttfamily 1906.07810}}.

\bibitem{Stieberger:2018edy}
S.~Stieberger and T.~R. Taylor, \emph{{Strings on Celestial Sphere}},
  \href{https://doi.org/10.1016/j.nuclphysb.2018.08.019}{\emph{Nucl. Phys.}
  {\bfseries B935} (2018) 388}
  [\href{https://arxiv.org/abs/1806.05688}{{\ttfamily 1806.05688}}].

\end{thebibliography}\endgroup
\bibliographystyle{JHEP}

\end{document}